\documentclass[12pt]{article}

\ifx\pdfoutput\undefined
\usepackage[dvips,bookmarks]{hyperref}
\else
\usepackage{hyperref}
\fi
\hypersetup{colorlinks=false,bookmarksopen,bookmarksnumbered,citecolor=blue,
   pdfstartview=FitH}

\usepackage{latexsym}
\usepackage{amssymb,amsfonts,amsmath}
\usepackage{graphicx} 
\usepackage{indentfirst}
\usepackage{bbm}
\usepackage{amssymb}
\usepackage{verbatim}
\usepackage{amsmath, amsthm,amssymb}
\usepackage{mathrsfs}
\usepackage{hyperref}
\usepackage{amsfonts}
\usepackage{dsfont}
\usepackage{slashed, tensor}
\usepackage{booktabs}
\usepackage{graphicx}
\usepackage{mathrsfs}

\parskip=\medskipamount

\usepackage[mathscr]{euscript} 

\newcommand {\cE}{{\cal E}}

\newcommand {\cG}{{\cal G}}
\newcommand {\cH}{{\cal H}}

\newcommand {\cJ}{{\cal J}}

\newcommand {\cN}{{\cal N}}

\def\a{\alpha}
\def\b{\beta}

\def\d{\delta}

\def\f{\phi}
\def\g{\gamma}
\def\G{\Gamma}

\def\j{\psi}
\def\k{\kappa}
\def\l{\lambda}
\def\m{\mu}

\def\o{\omega}

\def\q{\theta}
\def\r{\rho}
\def\s{\sigma}

\def\u{\upsilon}
\def\x{\xi}
\def\z{\zeta}
\def\D{\Delta}
\def\F{\Phi}
\def\J{\Psi}
\def\L{\Lambda}

\def\S{\Sigma}
\def\U{\Upsilon}

\def\ri{{\rm i}}
\def\re{{\rm e}}

\newcommand{\ad}{{\dot{\alpha}}}

\newcommand{\sSL}{\mathsf{SL}}

\newcommand{\sU}{\mathsf{U}}

\newcommand{\1}{{\underline{1}}}
\newcommand{\2}{{\underline{2}}}

\newcommand{\ve}{\varepsilon}

\newcommand{\ab}{{\a\b}}

\newcommand{\pa}{\partial}
\newcommand{\hf}{\frac12}

\newcommand{\vf}{\varphi}

\newcommand{\be}{\begin{equation}}
\newcommand{\ee}{\end{equation}}
\newcommand{\bea}{\begin{eqnarray}}
\newcommand{\eea}{\end{eqnarray}}
\newcommand{\non}{\nonumber}
\newcommand{\ba}{\begin{array}}
\newcommand{\ea}{\end{array}}

\newcommand{\bm}[1]{\mbox{\boldmath$#1$}}

\def\double #1{#1{\hbox{\kern-2pt $#1$}}}

\newcommand{\bsubeq}{\begin{subequations}}
\newcommand{\esubeq}{\end{subequations}}

\newcommand{\rd}{\mathrm d}
\newcommand{\floor}[1]{\left \lfloor #1 \right \rfloor}

\numberwithin{equation}{section}  

\newcommand{\pde}{\partial}
\newcommand{\mri}{\mathrm{i}}
\newcommand{\mrd}{\mathrm{d}}

\begin{document}

\begin{titlepage}
\begin{flushright}
March, 2016 \\
Revised version: July, 2016
\end{flushright}
\vspace{5mm}

\begin{center}
{\Large \bf 
Off-shell higher spin \mbox{$\bm{\cN=2}$} supermultiplets \\
 in three dimensions}
\\ 
\end{center}

\begin{center}

{\bf
Sergei M. Kuzenko and Daniel X. Ogburn
} \\
\vspace{5mm}

\footnotesize{
{\it School of Physics M013, The University of Western Australia\\
35 Stirling Highway, Crawley W.A. 6009, Australia}}  
~\\

\vspace{2mm}
~\\
\texttt{sergei.kuzenko@uwa.edu.au, daniel.ogburn@research.uwa.edu.au}\\
\vspace{2mm}

\end{center}

\begin{abstract}
\baselineskip=14pt

Off-shell higher spin $\cN=2$ supermultiplets in three spacetime dimensions 
(3D) are presented in this paper.
We propose gauge prepotentials for higher spin superconformal gravity 
and construct the corresponding gauge-invariant field strengths, which 
are proved to be conformal primary superfields.
These field strengths  
are higher spin generalisations of the (linearised)  $\cN=2$ super-Cotton tensor, 
which controls the superspace geometry of conformal supergravity.
We also construct the higher spin extensions of the linearised $\cN=2$ conformal  
supergravity action. 
We provide two dually equivalent off-shell formulations for massless
higher spin $\cN=2$ supermultiplets. They involve one and the same superconformal
prepotential but differ in the compensators used.
For the lowest superspin value 3/2,
these higher spin series terminate  at the linearised actions for 
the (1,1) minimal and $w=-1$ non-minimal $\cN=2$ Poincar\'e supergravity theories
constructed in arXiv:1109.0496. Similar to the pure 3D supergravity actions, 
their higher spin counterparts  propagate no  degrees of freedom.
However, the  massless higher spin supermultiplets are 
used to construct off-shell massive $\cN=2$ supermultiplets 
by combining the massless actions with those describing 
higher spin extensions of the linearised $\cN=2$ conformal supergravity. 
We also demonstrate that every higher spin super-Cotton tensor can 
be represented as a linear superposition of the equations of motion for the
corresponding massless higher spin supermultiplet,
with the coefficients being higher-derivative linear operators.

\end{abstract}

\vfill

\vfill
\end{titlepage}

\newpage
\renewcommand{\thefootnote}{\arabic{footnote}}
\setcounter{footnote}{0}

\tableofcontents


\allowdisplaybreaks

\section{Introduction}

In supersymmetric field theory,  it is of interest to construct off-shell 
supersymmetric extensions in diverse dimensions
of the (Fang-)Fronsdal actions 
for massless higher spin 
fields in Minkowski \cite{Fronsdal,FF} and anti-de Sitter
\cite{Fronsdal2,FF2} spacetimes.
In four spacetime dimensions (4D), this problem was solved in the early 1990s.
In the $\cN=1$ super-Poincar\'e case, 
the off-shell formulations
for massless higher spin supermultiplets
were developed in  \cite{KSP,KS93}. 
For each superspin\footnote{In four dimensions,
the massless 
multiplet of superspin $s$ describes two fields of spin $s$ and $s+\hf$; 
it is often denoted $(s,s+\hf)$.}
$s \geq 1$, 
half-integer \cite{KSP} and integer \cite{KS93},
these publications provided two dually equivalent off-shell
realisations in ${\cal N }= 1$ Minkowski superspace.
At the component level, each of the
two superspin-$s$ actions \cite{KSP,KS93} reduces, 
{\it upon} imposing a Wess-Zumino-type
gauge and eliminating the auxiliary fields,
to a sum of the spin-$s$ and  spin-$(s+1/2)$ actions \cite{Fronsdal,FF}.
The off-shell higher spin supermultiplets  of \cite{KSP,KS93} were generalised 
to the case of anti-de Sitter supersymmetry in \cite{KS94}.
Making use of the  $\cN=1$ supermultiplets
constructed in  \cite{KSP,KS93,KS94}, 
off-shell formulations for 4D $\cN=2$ massless
higher spin supermultiplets were presented in \cite{GKS1,GKS2}.\footnote{An 
important by-product of the higher spin construction given in \cite{GKS2} 
was the explicit description of the infinite dimensional superalgebra 
of Killing tensor superfields of 4D $\cN=1$ anti-de Sitter superspace.
This superalgebra corresponds to the rigid symmetries of the generating action for
the massless supermultiplets of arbitrary superspin
in 4D $\cN=1$ anti-de Sitter superspace, 
which was constructed in \cite{GKS2}. 
A generalisation of the concept of Killing tensor superfields given in \cite{GKS2} 
has recently been appeared in \cite{HL}.
}
A pedagogical review of the supersymmetric higher spin models 
proposed in \cite{KSP,KS93} is given in section 6.9 of \cite{BK}.\footnote{Section 6.9 
of \cite{BK} also contains a pedagogical review of 
the (Fang-)Fronsdal actions for free massless higher spin 
fields in 4D Minkowski space \cite{Fronsdal,FF}, including a  direct proof 
of the fact that that  the massless spin-$s$ action describes two helicity states $\pm s$. 
} 
A comprehensive review of the results of \cite{KSP,KS93,KS94}, 
including a detailed analysis of the component structure of the models constructed,
is given in \cite{Sibiryakov}.

In this paper, we present off-shell $\cN=2$ supersymmetric generalisations
of the 3D (Fang-)Fronsdal actions and derive their massive deformations. 
In principle, one may construct all 3D $\cN=2$ massless higher spin supermultiplets 
by applying an off-shell version of  
dimensional reduction $d=4 \to d=3$
to the 4D $\cN=1$ supermultiplets \cite{KSP,KS93}. 
Such a procedure has been carried out in \cite{KT-M11} to 
obtain one of the four off-shell actions (given in \cite{KT-M11})
for linearised 3D $\cN=2$ supergravity
(superspin $s={3}/{2}$). In practice, however, naive dimensional reduction 
is not quite efficient to deal with in the case of higher spin supermultiplets.
The point is that its application
to a 4D superspin-$s$ multiplet, with $s>3/2$,  leads to a superposition of several 
3D multiplets, one of which carries superspin $s$ and the others
correspond to lower superspin values.\footnote{In the case of a half-integer 
superspin $s=n+1/2$, with $n=2,3, \dots$, one of the 4D dynamical variables \cite{KSP} 
is a real unconstrained
superfield 
$H_{\a_1 \dots \a_n \ad_1 \dots \ad_n}= H_{(\a_1 \dots \a_n )( \ad_1 \dots \ad_n)}$.
Its dimensional reduction $d=4 \to d=3$ leads to a family of unconstrained symmetric 
superfields $H_{\a_1 \dots \a_{2n}}$, $H_{\a_1 \dots \a_{2n-2}}$, $\cdots$, $H$,
of which only $H_{\a_1 \dots \a_{2n}}$ is required to describe a massless 3D 
supermultiplet.
In the supergravity case, $s=3/2$, 
dimensional reduction $d=4 \to d=3$ leads to two multiplets, 
an off-shell $\cN=2$ supergravity multiplet and an Abelian vector multiplet 
 \cite{KT-M11}.} 
Some work is required in order to disentangle the superspin-$s$ multiplet
from the lower-superspin ones, which is actually quite nontrivial.  
It proves to be  more efficient to recast
the 4D gauge principle of  \cite{KSP,KS93} in a 3D form and use it to construct 
gauge-invariant actions. This is our approach in the present paper. 

In three dimensions, the massless spin-$s$  actions of \cite{Fronsdal,FF} 
are known to propagate no local degrees of freedoms for $s>1$.\footnote{See
Appendix \ref{AppendixB} for a direct proof.}
Of course, this is consistent with the fact that the notion of 3D spin is 
well defined only in the massive case \cite{Binegar}.
When speaking of a 3D massless spin-$s$ theory, we will refer to the kinematic 
structure of the field variables, 
their gauge transformation laws and the gauge-invariant action. 
One reason to study such a theory is that it may be deformed 
(say, by including auxiliary lower-spin fields and adding mass terms) 
to result in a model describing a massive spin-$s$ field. 

There have appeared two different constructions of Lagrangian models for 3D massive higher spin 
fields \cite{TV,BSZ-non}. The approach of \cite{BSZ-non} has been used 
to formulate {\it on-shell} models for massive $\cN=1$ higher spin supermultiplets
\cite{BSZ}. In this paper we will pursue an alternative approach to address the problem
of constructing {\it off-shell} massive $\cN=2$ higher spin supermultiplets.
Our approach will be based on deriving a higher spin generalisation of the $\cN=2$ 
super-Cotton tensor\footnote{Upon fixing the super-Weyl and local $\sU(1)_R$ symmetries,
 the super-Cotton tensor derived in \cite{K12} reduces to that introduced earlier 
 by Zupnik and Pak  \cite{ZupnikPak}.} 
\cite{K12,BKNT-M} that can be used to write down
a topological mass term. 

An important feature of 3D gauge theories 
is the possibility to generate the mass 
for gauge fields of different spin
by adding to the massless action 
a gauge-invariant Chern-Simons-type term of topological origin.
This idea has been used to construct 
topologically massive electrodynamics \cite{Siegel,Schonfeld,DJT}, 
topologically massive gravity \cite{DJT} and topologically massive 
$\cN=1$ supergravity \cite{DK,Deser}. The latter theory admits generalisations 
with $\cN>1$, including the off-shell topologically massive supergravity theories with 
$\cN=2$ \cite{KLRST-M} and $\cN=3$ and $\cN=4$ \cite{KN14}. 
In the case of 3D supergravity theories, the topological mass term
may be interpreted as an 
 action for conformal supergravity (see \cite{FT} for a review of 
4D conformal supergravity theories).
The off-shell actions for $\cN$-extended conformal supergravity theories 
were constructed in \cite{vanN85} for $\cN=1$, 
\cite{RvanN86} for $\cN=2$, \cite{BKNT-M2} for $\cN=3,4,5$, 
and \cite{NT,KNT-M13} for $\cN=6$.\footnote{The  component actions
for $\cN=1,2$  conformal supergravities \cite{vanN85,RvanN86} 
have been re-derived within the universal superspace setting of \cite{BKNT-M2}.} 
An arbitrary variation of such an action with respect to a supergravity prepotential 
is given in terms of the $\cN$-extended super-Cotton tensor \cite{BKNT-M}. 
This means that a linearised supergravity action is determined by the linearised
super-Cotton tensor, $W(H)$. The corresponding Lagrangian is symbolically
$L_{\rm CSG}= H  \cdot W(H)$, where $H$ is  the linearised conformal supergravity 
prepotential. The super-Cotton tensor $W(H)$ is a unique field strength 
being invariant under the linearised gauge transformations of conformal supergravity.

Our construction of the linearised higher spin superconformal actions 
is analogous to that of the 3D   higher spin conformal gravity actions derived by Pope and Townsend
\cite{PopeTownsend} (the 3D analogues of the conformal higher spin actions
pioneered by Fradkin and Tseytlin \cite{FT}).
The Pope-Townsend conformal action for the  spin-$s$ field makes use of the linearised 
spin-$s$ Cotton tensor (which can be read off from the action (31) in \cite{PopeTownsend}).
The $s=3$ case was studied earlier in \cite{DD}. For recent discussions of 
the linearised higher spin Cotton tensors \cite{PopeTownsend} and their generalisations,
see \cite{HHL,LN} and references therein.

This paper is organised as follows. Section 2 is devoted to general properties 
of transverse and longitudinal linear superfields. Section 3 is concerned with 
on-shell massive fields and $\cN=2$ superfields. 
Two series of off-shell actions for massless half-integer superspin multiplets are introduced in 
section 4. Section 5 is devoted to a brief discussion 
of the component reduction of the models presented in section 4. 
In section 6 we present $\cN=2$ superconformal higher spin actions 
and derive a higher spin extension of the linearised $\cN=2$ super-Cotton tensor.
Off-shell actions for massive higher spin supermultiplets 
are presented in section 7. Concluding comments are given in section 8.
The main body of the paper is accompanied by four appendices. 
Appendix A summarises our notation and conventions.
Appendix B is devoted to the 3D  (Fang-)Fronsdal massless actions 
in the two-component notation. 
The component structure of the massless 
superspin-$(s+\hf)$  model  \eqref{eq:FlatAction3DTrans} 
is studied in Appendix C.
Appendix D is devoted to the proof of two fundamental 
properties of the superconformal field strength  \eqref{eq:HSFSUniversal}.


\section{Linear superfields}\label{section2}

A symmetric rank-$n$ spinor superfield, $\G_{\a_1 \cdots \a_n} = \G_{(\a_1 \cdots \a_n)}$,
is called {\it transverse linear} if it obeys the constraint
\begin{subequations} 
\bea
\bar D^\b \G_{\b \a_1 \dots \a_{n-1}} =0~, \qquad n>0~.
\label{TL}
\eea
A symmetric rank-$n$ spinor superfield, $G_{\a_1 \cdots \a_n} = G_{(\a_1 \cdots \a_n)}$,
is called {\it longitudinal linear} if it obeys the constraint
\bea
\bar D_{(\a_1} G_{\a_2 \dots \a_{n+1})} =0~,
\label{LL}
\eea
\end{subequations}
which for $n=0$ is equivalent to the chirality condition
\bea
\bar D_\a G=0~.
\eea
The constraints \eqref{TL} and \eqref{LL} imply that  
$\G_{\a_1 \cdots \a_n}$ and $G_{\a_1 \cdots \a_n}$
are linear superfields in the usual sense:
\begin{subequations}
\bea
\bar D^2 \G_{\a_1 \cdots \a_n} &=& 0~, \label{ordinary_linear}\\ 
\qquad \bar D^2 G_{\a_1 \cdots \a_n} &= &0~.
\eea
\end{subequations}
In the case $n=0$, the transverse constraint \eqref{TL} is not defined, 
but its corollary \eqref{ordinary_linear} can be used.
In four dimensions, the transverse
and longitudinal linear superfields were introduced for the first time by Ivanov and Sorin
\cite{IS} (who built on the earlier  results by Salam and Strathdee \cite{SS} and Sokatchev  \cite{Sokatchev} in the super-Poincar\'e case)
as a means to realise the irreducible representations of 
the $\cN=1$ anti-de Sitter supersymmetry. As dynamical variables,
such superfields were used for the first time in \cite{KSP,KS93,KS94}.

We assume that $\G_{\a_1 \cdots \a_n}$ and $G_{\a_1 \cdots \a_n}$ are complex
and the differential conditions \eqref{TL} and \eqref{LL} are the only constraints
these superfields obey. 
The constraints \eqref{TL} and \eqref{LL} can be solved in terms of complex 
unconstrained prepotentials $\x_{\a_1 \dots \a_{n+1} } =\x_{(\a_1 \dots \a_{n+1} )}$ 
and $\z_{\a_1 \dots \a_{n-1}} = \z_{(\a_1 \dots \a_{n-1} )} $ according to the rules 
\begin{subequations} \label{prepotentials}
\bea
\G_{\a_1 \dots \a_n } &=& \bar D^\b \x_{\b \a_1 \dots \a_n } \ , \\ 
G_{\a_1 \dots \a_n} &=& \bar D_{(\a_1} \z_{\a_2 \dots \a_n)} \ . 
\eea
\end{subequations}
There is a natural arbitrariness in the choice of the prepotentials $\x$ and $\z$, namely,
\begin{subequations}
\bea
\d \x_{\a_1 \dots \a_{n+1}} &=& \G_{\a_1 \dots \a_{n+1}} \ , \\
\d \z_{\a_1 \dots \a_{n-1} } &=& G_{\a_1 \dots \a_{n-1} } \ .
\eea
\end{subequations}
Here, the gauge parameter $\G_{\a(n+1)}$ is a transverse linear superfield, 
and $G_{\a(n-1)}$ is a longitudinal linear one. 
As a result, there emerge the transverse and longitudinal gauge hierarchies:
\begin{subequations}
\bea
&& \G_{\a(n)}  \longrightarrow \G_{\a(n+1)}  \longrightarrow \G_{\a(n+2)}\dots~, \\
&& G_{\a(n)}  \longrightarrow G_{\a(n-1)}  \longrightarrow G_{\a(n-2)}
  \dots  \longrightarrow G \ .
\eea 
\end{subequations}
Thus, in accordance with the terminology of gauge theories with linearly 
dependent generators \cite{BV}, any Lagrangian theory described by a transverse 
(longitudinal)  linear superfield $\G_{\a_1 \dots \a_n }$ 
($G_{\a_1 \dots \a_n }$) can be considered as the theory of an unconstrained
prepotential $\x_{\a_1 \dots \a_{n+1} }$ 
($\z_{\a_1 \dots \a_{n-1} } $) with an additional gauge invariance of an infinite 
(finite) stage of reducibility. 

Suppose we are given a supersymmetric field theory described by 
a transverse linear superfield $\G_{\a(n)}$ and its conjugate $\bar \G_{\a(n)}$, for $n>0$,
with an action functional $S[\G, \bar \G]$. Such a theory possesses a dual formulation, 
$S_{\rm D}[G,\bar G]$, described in terms of a longitudinal linear superfield 
$G_{\a(n)}$ and its conjugate $\bar G_{\a(n)}$. The latter theory is obtained by introducing 
a first-oder action of the form
\bea
S[V, \bar V , G, \bar G]= S[V, \bar V] + 
\int {\rm d}^3x {\rm d}^2 \q \rd^2 \bar \q \, 
\Big( V^{\a(n)} G_{\a(n)} + (-1)^n\bar V^{\a(n)} \bar G_{\a(n)} \Big)~,
\label{first-order-action}
\eea
where the symmetric rank-$n$ spinor $V_{\a(n)}$ is a complex unconstrained superfield. 
The first term in the action, $S[V, \bar V]$, is obtained from $S[\G, \bar \G]$
by the replacement $\G_{\a(n)} \to V_{\a(n)}$. Varying \eqref{first-order-action}
with respect to $G_{\a(n)}$ gives $V_{\a(n)} = \G_{\a(n)}$, and then the second
term in \eqref{first-order-action} drops out, due to the identity 
\bea
 \int {\rm d}^3x {\rm d}^2 \q \rd^2 \bar \q \, 
\G^{\a(n)} G_{\a(n)} =0~.
\eea
As a result, the first-order action 
reduces to the original one, $S[\G, \bar \G]$. On the other hand, we can 
consider the equation of motion for $V^{\a(n)}$, 
\bea
\frac{\d }{\d V^{\a(n)} } S[V, \bar V] + G_{\a(n)} =0~,
\eea
and the conjugate equation.
We assume that these equations are uniquely solved to give $V_{\a(n)}$ 
as a functional of $G_{\a(n)}$ and $\bar G_{\a(n)}$. Substituting this solution 
back into  \eqref{first-order-action}, we end up with the dual action 
$S_{\rm D}[G,\bar G]$. 

A real transverse linear superfield $T_{\a_1 \cdots \a_n} = T_{(\a_1 \cdots \a_n)}$ 
is characterised by the properties 
\bea
\bar T_{\a_1 \cdots \a_n} =  T_{\a_1 \cdots \a_n} ~,\qquad 
\bar D^\b T_{\b \a_1 \cdots \a_{n-1}} = 0\quad \Longleftrightarrow \quad
D^\b T_{\b \a_1 \cdots \a_{n-1}} = 0 ~ .
\label{RTL}
\eea
The second-order differential operator
\bea
\D = \frac{\ri}{2} D^\a \bar D_\a
\label{Delta}
\eea
acts on the space of such superfields. Indeed, 
 $\D T_{\a_1 \cdots \a_n}$ is real and one may check that 
\bea 
\bar D^\b \D T_{ \b \a_1 \cdots \a_{n-1} } =0~, \qquad 
 D^\b \D T_{\b \a_1 \cdots \a_{n-1}} 
= 0 \ .
\eea


\section{Massive (super)fields}

In this section we discuss on-shell (super)fields which realise
the massive representations of the (super-)Poincar\'e group. 

\subsection{Massive fields}

Let $P_a$ and $J_{ab}= -J_{ba}$ be the generators of the 3D Poincar\'e group.
The Pauli-Lubanski scalar 
\bea
W:= \hf \ve^{abc}P_a J_{bc} = -\hf P^{\a\b} J_{\a\b}
\label{PauliL}
\eea
commutes with the generators $P_a$ and $J_{ab}$.
Irreducible unitary representations of the Poincar\'e group 
are labelled by two parameters,  mass $m$ and helicity $\l$, 
which are associated with the Casimir operators, 
\bea
 P^a P_a = -m^2 {\mathbbm 1} ~, \qquad W=m \l {\mathbbm 1}~.
 \label{Casimirs}
 \eea
 One defines $|\l|$ to be the spin. 
 
In the case of field representations, 
\bea
W= \hf \pa^{\ab} M_{\a\b}~,
\eea
where the action of $M_{\a\b}=M_{\b\a}$ on a field 
$\f_{\g_1 \cdots \g_n} = \f_{(\g_1 \cdots \g_n)}$ is defined by 
\bea
M_{\a\b} \f_{\g_1 \cdots \g_n} = \sum_{i=1}^n
\ve_{\g_i (\a} \f_{\b) \g_1 \cdots \widehat{\g_i} \dots\g_n}~,
\eea
where the hatted index of $\f_{\b \g_1 \cdots \widehat{\g_i} \dots\g_n}$  is omitted.

For $n>1$, a massive field, $\f_{\a_1 \cdots \a_n} 
= \bar \f_{\a_1\dots \a_n} = \f_{(\a_1 \cdots \a_n)}  $,
is a real symmetric rank-$n$ spinor field
which obeys the differential conditions \cite{TV} (see also \cite{BHT})
\begin{subequations}
\bea
\pa^{\b\g} \f_{\b\g\a_1 \cdots \a_{n-2}} &=&0~, \label{dif_sub} \\
\pa^\b{}_{(\a_1} \f_{\a_2 \dots \a_n)\b} &=& m \s \f_{\a_1\dots \a_n}~,
\qquad \s =\pm 1~.
\label{mass3.6}
\eea 
\end{subequations}
In the spinor case, $n=1$, eq.  \eqref{dif_sub} is absent, and it is the Dirac equation 
\eqref{mass3.6} which defines a massive field. 
It is easy to see that  \eqref{dif_sub} and  \eqref{mass3.6}
imply the mass-shell equation\footnote{The equations \eqref{dif_sub} and 
\eqref{mass-shell} proves to be equivalent to the 3D Fierz-Pauli field equations \cite{FP}.}
\bea
(\Box -m^2 ) \f_{\a_1 \cdots \a_n} =0~,
\label{mass-shell}
\eea
which is the first equation in \eqref{Casimirs}. In the case $n=1$,
eq. \eqref{mass-shell} follows from the Dirac equation \eqref{mass3.6}. 
The second relation in \eqref{Casimirs} also holds, with 
\bea
\l = \frac{n}{2} \s~.
\eea 

\subsection{Massive superfields}\label{subsection3.2}

Let $P_{a}$,  $J_{ab}= -J_{ba}$, $Q_\a$ and $\bar Q_\a$
 be the generators of the 3D $\cN=2$ super-Poincar\'e group.
 The supersymmetric extension of the Pauli-Lubanski scalar 
 \eqref{PauliL} is the following operator  \cite{MT}
\bea
Z= W-\frac{\ri}{4} Q^\a \bar Q_\a 
= \hf \ve^{abc}P_a J_{bc} -\frac{\ri}{4} Q^\a \bar Q_\a ~,
\label{super-helicity}
\eea
which commutes with the supercharges,
\bea
[Z,Q_\a ] =[Z,\bar Q_\a] =0~.
\eea
Irreducible unitary representations of the super-Poincar\'e group 
are labelled by two parameters,  mass $m$ and superhelicity $\k$, 
which are associated with the Casimir operators, 
\bea
 P^a P_a = -m^2 {\mathbbm 1} ~, \qquad Z=m \k {\mathbbm 1}~.
 \label{Casimirs-super}
 \eea
 Our definition of the superhelicity agrees with \cite{MT}.
 It is instructive to compare the operator $Z$, eq. 
 \eqref{super-helicity},  with the 4D $\cN=1$  superhelicity operator
 introduced in  \cite{BK}.
 The massive representation of superhelicity $\k$ is a direct sum of 
 four massive representations
 of the Poincar\'e group with helicity values $(\k-\hf, \k, \k, \k+\hf)$. 
 The parameter  $|k|$ is referred to as superspin \cite{MT}.
  
In the case of superfield representations, the superhelicity operator
may be expressed in the following manifestly supersymmetric form 
\bea
Z= \hf \pa^{\ab} M_{\a\b}+\hf \D~,
\label{super-helicity2}
\eea
where the operator $\D$ is given by \eqref{Delta}.

For $n>0$, a massive superfield, 
$\cE_{\a_1 \cdots \a_n} 
= \bar \cE_{\a_1\dots \a_n} = \cE_{(\a_1 \cdots \a_n)}  $,
is a real symmetric rank-$n$ spinor 
which obeys the differential conditions \cite{KNT-M}
\begin{subequations}\label{313}
\bea
\bar D^\b \cE_{\b \a_1 \cdots \a_{n-1}} &=&
D^\b \cE_{\b \a_1 \cdots \a_{n-1}} = 0 \quad \Longrightarrow \quad
\pa^{\b\g} \cE_{\b\g\a_1\dots \a_{n-2}} =0
~ , \label{313a} \\
\D \cE_{\a_1 \dots \a_n} &=& m \s \cE_{\a_1 \dots \a_n}~, \qquad \s =\pm 1~.
\eea
\end{subequations}
Due to the identity
\bea
\Box = \D^2 +\frac{1}{16} \left\{ D^2 , \bar D^2 \right\}~,
\eea
eqs. \eqref{313} lead to the mass-shell equation
\bea
(\Box -m^2) \cE_{\a_1 \dots \a_n} =0~.
\eea
One may also check that 
\bea
\D \cE_{\a_1 \dots \a_n}  = \pa^\b{}_{(\a_1} \cE_{\a_2 \dots \a_n)\b} ~,
\eea
as a consequence of \eqref{313a}.
We conclude that $\cE_{\a_1 \dots \a_n} $
is an eigenvector of the superhelicity operator 
\eqref{super-helicity2}, 
\bea
Z \cE_{\a_1 \dots \a_n}  = m \k \cE_{\a_1 \dots \a_n}  ~, 
\qquad \k = \hf (n+1)  \s  ~.
\eea

For completeness, we also consider massive scalar superfields. 
The massive multiplet of superhelicity $\k_\s =\pm \hf \equiv  \hf \s$
is described by a real scalar superfield   $\cG_\s = \bar \cG_\s$, which is
constrained by 
\bea
\D \cG_\s = m \s \cG_\s ~, 
\label{3.18}
\eea
with $m>0$ the mass parameter. This equation implies that $\cG_\s$ is linear, 
$\bar D^2 \cG_\s= D^2 \cG_\s=0$. 
It follows from \eqref{3.18} that $Z \cG_\s =  m \k_\s \cG_\s$.

Constraint  \eqref{3.18} is the equation of motion 
for a supersymmetric Chern-Simons theory with action
\bea
S_{\rm CS}[V] =-\hf \int {\rm d}^3x {\rm d}^2 \q \rd^2 \bar \q \,\Big\{ 
\cG^2 - m \s V \cG \Big\}~, \qquad \cG := \D V~,
\eea
with $V$ the gauge prepotential of the vector multiplet. 

The superhelicity $\k =0$ multiplet is described by a chiral superfield $\F$, 
$\bar D_\a \F =0$,  constrained by 
\bea
-\frac{1}{4} D^2 \F +m \bar \F =0~.
\eea
This is the equation of motion for the model 
\bea
S[\F, \bar \F] =\int {\rm d}^3x {\rm d}^2 \q \rd^2 \bar \q \,\bar \F \F + \frac{m}{2} \Big\{ 
\int {\rm d}^3x {\rm d}^2 \q\, \F^2 +{\rm c.c.}\Big\}~.
\eea


\section{Massless half-integer superspin multiplets}

We fix an integer $s >1$ and consider two sets of superfield dynamical variables: 
\begin{align}
\mathcal{V}^{\perp}=&\big\{H_{\a(2s)},\,\Gamma_{\a(2s-2)},
\,\bar{\Gamma}_{\a(2s-2)} \big\} ~;& 
\label{TL4.1}\\
\mathcal{V}^{\parallel}=&\big\{H_{\a(2s)},\,G_{\a(2s-2)},\,\bar{G}_{\a(2s-2)}
\big\}~.
&
\label{LL4.2}
\end{align}
In both case 
${H}_{\a(2s)} = H_{(\a_{1}\a_{2}....\a_{2s-1}\a_{2s})}$ is an unconstrained 
real superfield. The complex superfields $\G_{\a(2s-2) }= \G_{(\a_1 \dots\a_{2s-2} )}$ and $G_{\a(2s-2)}= G_{(\a_1 \dots \a_{2s-2})}$ 
are  transverse and longitudinal, respectively.

We postulate the following linearised gauge transformations
for the dynamical superfields introduced
\begin{align}
\delta H_{\a(2s)}=& g_{\a(2s)}+\bar{g}_{\a(2s)}\nonumber &\\
\equiv&\bar{D}_{(\a_{1}}L_{\a_{2}...\a_{2s})}-D_{(\a_{1}}\bar{L}_{\a_{2}...\a_{2s})} ~,\label{eq:GTransHFlat} &\\
\delta \Gamma_{\a(2s-2)}=&\frac{s}{2s+1}\bar{D}^{\b_{1}}D^{\b_{2}}\bar{g}_{\a(2s-2)\b(2)}\nonumber &\\
=&-\frac{1}{4}\bar{D}^{\b}D^{2}\bar{L}_{\b\a(2s-2)} ~,
\label{eq:GTransGammaFlat} 
&\\
\delta G_{\a(2s-2)}=&\frac{s}{2s+1}D^{\b_{1}}\bar{D}^{\b_{2}}g_{\b(2)\a(2s-2)}+\ri s\pde^{\b(2)}g_{\b(2)\a(2s-2)}\nonumber &\\
=&-\frac{1}{4}\bar{D}^{2}D^{\b}L_{\b\a(2s-2)}+\mri(s-1)\pde^{\b_{1}\b_{2}}\bar{D}_{(\a_{1}}L_{\a_{2}...\a_{2s-2})\b_{1}\b_{2}}~. 
\label{eq:GTransGFlat} &
\end{align}
Here the complex gauge parameter  $g_{\a(2s)} =g_{(\a_1 \dots \a_{2s})} $ 
is an arbitrary longitudinal linear superfield. 
 It can be expressed 
 in terms of an unconstrained complex parameter 
 $L_{\a(2s-1)} =L_{(\a_1 \dots \a_{2s-1})} $ by the rule
\begin{equation}
g_{\a(2s)}=\bar{D}_{(\a_{1}}L_{\a_{2}...\a_{2s})}~.
\label{g4.6}
\end{equation}
Since the gauge transformations are linearised, only the $L_{\alpha(2s-1)}$ and $\bar{L}_{\alpha(2s-1)}$ fermionic gauge parameter superfields can appear in the gauge transformation.

The two sets of dynamical variables, 
$\mathcal{V}^{\perp}$ and $\mathcal{V}^{\parallel}$, 
give rise to two gauge-invariant actions, transverse and longitudinal ones,  
which are dual to each other.

Let us introduce unconstrained prepotentials, $\x_{\a(2s-1)}$
and $\z_{\a (2s-3)}$, 
for the constrained superfields $\G_{\a(2s-2) }$ and $G_{\a(2s-2)}$ 
according to the rule \eqref{prepotentials}.
The gauge transformations of  $\G_{\a(2s-2) }$ and $G_{\a(2s-2)}$ 
are induced by the following variations of the 
prepotentials 
\bea
\d \x_{\a(2s-1)} &=& -\frac{1}{4}D^{2}\bar{L}_{\a(2s-1)}~, \\
\d \z_{\a (2s-3) } &=& -\hf \bar D^\b D^\g L_{\a(2s-3) \b \g} 
+\ri (s-1) \pa^{\b\g} L_{\a(2s-3) \b \g}~. 
\eea

In what follows, 
we will use the notation 
$ \mrd^{3|4}z= \mrd^{3}x \mrd^{2}\theta \mrd^{2}\bar{\theta}$
for the full superspace measure. 


\subsection{Transverse formulation}

The transverse formulation for a massless superspin-$(s+\hf)$ multiplet is
described by the action 
\begin{align}
S^{\perp}_{s+\frac{1}{2}}[H,\Gamma, \bar \G]
=& \Big(-\frac{1}{2}\Big)^{s}\int \mrd^{3|4}z\,
\bigg\{\frac{1}{8}H^{\a(2s)}D^{\b}\bar{D}^{2}D_{\b}H_{\a(2s)}
\nonumber &\\
&+H^{\a(2s)}\left(D_{\a_1}\bar{D}_{\a_2}\Gamma_{\a_3 \dots \a_{2s}}
-\bar{D}_{\a_1} D_{\a_2}\bar{\Gamma}_{\a_3\dots \a_{2s}}\right)\nonumber &\\
&+\frac{2s-1}{s}\bar{\Gamma}\cdot \Gamma + \frac{2s+1}{2s}\left(\Gamma \cdot \G+\bar{\Gamma}\cdot \bar \G
\right)\bigg\}~.
\label{eq:FlatAction3DTrans}
&
\end{align}
It may be shown that the action 
is invariant under the  gauge transformations \eqref{eq:GTransHFlat} 
and \eqref{eq:GTransGammaFlat}. The requirement of gauge invariance  
fixes this action uniquely  up to an constant.

Consider an arbitrary variation of the action 
\bea
\d S^{\perp}_{s+\frac{1}{2}}
= \Big(-\frac{1}{2}\Big)^{s} \int \mrd^{3|4}z\,\Big\{ \d H^{\a(2s)} E^{\perp}_{\a(2s)}
-\d \x^{\a(2s-1)} F_{\a(2s-1)} 
-\d \bar \x_{\a(2s-1)} \bar F^{\a(2s-1)} \Big\}~,~~~~~
\eea
where we have introduced 
the gauge-invariant field strengths
\begin{subequations} \label{4.11ab}
\bea
E^{\perp}_{\a(2s)}
&=&\frac{1}{4}D^{\b}\bar{D}^{2}D_{\b}H_{\a(2s)}
+D_{(\a_{1}}\bar{D}_{\a_{2}}\Gamma_{\a_3 \dots \a_{2s})}
-\bar{D}_{(\a_1}D_{\a_{2}}\bar{\Gamma}_{\a_3 \dots \a_{2s})}~,~~~\\
F_{\a(2s-1)}
&=&-\frac{1}{2}\bar{D}^{2}D^{\b}H_{\a(2s-1)\b}
+\frac{2s-1}{s}\bar{D}_{(\a_{1}}\bar{\Gamma}_{\a_2 \dots \a_{2s-1})} \non \\
&&
+\frac{2s+1}{s}\bar{D}_{(\a_{1}}\Gamma_{\a_2 \dots \a_{2s-1})}~.
\eea
\end{subequations}
The field strengths are related to each other by the Bianchi identity
\begin{equation}
D^{\b}E^\perp_{\b\a(2s-1)}=\frac{1}{4}D^{2}F_{\a(2s-1)} \label{eq:HSBianchiIdentity1},
\end{equation}
The equations of motion for the theory are
\begin{equation}
E^{\perp}_{\a(2s)}=0, \qquad F_{\a(2s-1)}=0
 \label{eq:HSFSMasslessEOM}~. 
\end{equation}


\subsection{Longitudinal formulation}

In accordance with our general discussion in section \ref{section2}, the theory 
with action \eqref{eq:FlatAction3DTrans} possesses a dual formulation. 
It is obtained by considering the following first-order action
\bea
S[H,G, \bar G,V, \bar V]:&=& S^{\perp}_{s+\frac{1}{2}}[H,V,\bar V]
-\frac{2}{s}
\Big(-\frac{1}{2}\Big)^{s}\int \mrd^{3|4}z\,
\Big\{G \cdot V +\bar{G}\cdot \bar{V}\Big\} \non \\
&=& \left(-\frac{1}{2}\right)^{s}\int \mrd^{3|4}z\,
\bigg\{\frac{1}{8}H^{\a(2s)}D^{\gamma}\bar{D}^{2}D_{\gamma}H_{\a(2s)}\nonumber \\
&&+H^{\a(2s)}\left(D_{\a_1}\bar{D}_{\a_2}V_{\a_3 \dots \a_{2s}}
-\bar{D}_{\a_{1}} D_{\a_{2}}\bar{V}_{\a_3 \dots \a_{2s}}\right)\nonumber \\
&&+\frac{2s-1}{s}\bar{V}\cdot V + \frac{2s+1}{2s}\left(V^{2}+\bar{V}^{2}\right)
-\frac{2}{s}(G\cdot V+\bar{G}\cdot \bar{V} )\bigg\}~.~~~~~
\label{eq:FlatAction3DAuxiliary}
\eea
Here $V_{\a(2s-2)}$ is an unconstrained  complex superfield, 
while the Lagrange multiplier $G_{\a(2s-2)}$ is constrained to be 
a complex longitudinal linear superfield.
With the normalisation of the Lagrange multiplier chosen, the action 
\eqref{eq:FlatAction3DAuxiliary} proves to be invariant under the gauge transformations
\eqref{eq:GTransHFlat} and \eqref{eq:GTransGFlat} accompanied by
\begin{align}
\delta V_{\a(2s-2)}
=- \frac{1}{4}\bar{D}^{\b}D^{2}\bar{L}_{\b\a(2s-2)}~.&
\end{align}

Varying \eqref{eq:FlatAction3DAuxiliary} with respect to $G_{\a(2s-2)}$ gives
$V_{\a(2s-2)}=\G_{\a(2s-2)}$, and then \eqref{eq:FlatAction3DAuxiliary} reduces 
to the transverse action \eqref{eq:FlatAction3DTrans}. On the other hand, we can 
first consider the equation of motion for $V_{\a(2s-2)}$ and its conjugate, 
which imply
\bea
V_{\a(2s-2)}&=&-\frac{1}{8}[D^\b,\bar{D}^{\g}]H_{\b \g \a_1 \dots \a_{2s-2}}
-\frac{\ri}{2}s\pde^{\b \g}H_{\b \g\a_1 \dots \a_{2s-2}} \non \\
& &+\frac{2s+1}{4s}G_{\a(2s-2)}-\frac{2s-1}{4s}\bar{G}_{\a(2s-2)}~.
\eea
Using this and the conjugate relation, we can express the action 
 \eqref{eq:FlatAction3DAuxiliary} in terms of the dynamical variables
 $H_{\a(2s)}$, $G_{\a(2s-2)} $ and  $\bar G_{\a(2s-2)} $.
 The result is 
\begin{align}
S^{\parallel}_{s+\frac{1}{2}}[H,G, \bar G]=& \left(-\frac{1}{2}\right)^{s} \int \mrd^{3|4}z\,
\bigg\{\frac{1}{8}H^{\a(2s)}D^{\gamma}\bar{D}^{2}D_{\gamma}H_{\a(2s)}
\nonumber &\\
&-\frac{1}{16}\Big( [D_{\b_{1}},\bar{D}_{\b_{2}}]H^{\b_1 \b_2\a(2s-2)}\Big)
[D^{\gamma_{1}},\bar{D}^{\gamma_{2}}]H_{\gamma_1 \g_2\a(2s-2)}\nonumber &\\
&+\frac{s}{2} \Big(\pde_{\b_{1}\b_{2}}H^{\b_1\b_2\a(2s-2)}\Big)
\pde^{\gamma_{1}\gamma_{2}}H_{\gamma_1 \g_2\a(2s-2)}\nonumber &\\
&+\mri\frac{2s-1}{2s}\left(G-\bar{G}\right)^{\a(2s-2)}
\pde^{\b_1\b_2}H_{\b_1\b_2\a(2s-2)}\nonumber &\\
&+
\frac{2s-1}{2s^2}G\cdot \bar{G}
-
\frac{2s+1}{4s^2}\left(G\cdot G+\bar{G} \cdot \bar G\right)\bigg\}~.\label{eq:FlatAction3DLongitudinal}
&
\end{align}
This action is invariant under the gauge transformations \eqref{eq:GTransHFlat} and \eqref{eq:GTransGFlat}. It defines the longitudinal formulation of the theory.
By construction, the transverse and longitudinal formulations, \eqref{eq:FlatAction3DTrans} and \eqref{eq:FlatAction3DLongitudinal}, are dual to each other.

Computing the first variational derivatives of the action \eqref{eq:FlatAction3DLongitudinal} with respect to the prepotentials, 
we obtain the following gauge-invariant field strengths 
\bea
E^{\parallel}_{\a(2s)}:
&=&\frac{1}{4}D^{\b}\bar{D}^{2}D_{\b}H_{\a(2s)}
-\frac{1}{8} [D_{(\a_1}, \bar D_{\a_2}]
[D^{\b_{1}},\bar{D}^{\b_{2}}]
H_{\a_3 \dots \a_{2s})\b_1 \b_2}\non \\
&&-s \pde^{\b(2)}\pde_{(\a_1\a_2}H_{\a_3 \dots \a_{2s})\b(2)}
-\ri\frac{2s-1}{2s}\pde_{(\a_1\a_2}(G-\bar{G})_{\a_3 \dots \a_{2s})}~,
\nonumber \\
B_{\a(2s-3)}:
&=&-\mri\frac{2s-1}{2s}\bar{D}^{\gamma}\pde^{\b(2)}H_{\a(2s-3)\b(2)\gamma}+\frac{1}{s}\frac{2s-1}{2s}\bar{D}^{\gamma}(G-\bar{G})_{\gamma\a(2s-3)}~.
\label{eq:HSFunctionalDerivativesLongitudinal}
\eea
They are related to each other by the  Bianchi identity
\begin{align}
\bar{D}^{\b}E^{\parallel}_{\b\a(2s-1)}=&\frac{1}{2}D_{(\a_1}\bar{D}_{\a_2}
B_{\a_3 \dots \a_{2s-1})}-\mri (s-1)\pde_{(\a_1\a_2}B_{\a_3 \dots \a_{2s-1})}~.&
\label{lon4.19}
\end{align}
The equations of motion are
\begin{equation}
E^{\parallel}_{\a(2s)}=0~, \qquad B_{\a(2s-3)}=0~.
\label{lon4.20}
\end{equation}

\subsection{Linearised supergravity models}

For the case $s=1$, the longitudinl action
\eqref{eq:FlatAction3DLongitudinal} takes the form 
\bea
S^{\parallel}_{{3/2}}[H,G, \bar G]= -\hf \int \mrd^{3|4}z\,
\bigg\{\frac{1}{8}H^{\a\b}D^{\gamma}\bar{D}^{2}D_{\gamma}H_{\a\b}
-\frac{1}{16}
([D_\a,\bar{D}_\b]H^{\a\b})^2 &&
\nonumber \\
+\frac{1}{2}
(\pde_{\a\b}H^{\a\b})^2
+\frac{\ri}{2} (G-\bar{G}) \pde_{\a\b}H^{\a\b}
+\hf
G \bar{G}
\bigg\}~,&&
\label{eq:FlatAction3DLongitudinal1}
\eea
where the compensator $G$ is  chiral, $\bar D_\a G=0$.
This action proves to coincide with the linearised action for type I supergravity
\cite{KT-M11} upon rescaling $3G = \s$. 
The action is invariant under the gauge transformations 
\begin{subequations}
\bea
\delta H_{\a \b}&=& g_{\a\b}+\bar{g}_{\a\b}
=\bar{D}_{(\a}L_{\b )}-D_{(\a}\bar{L}_{\b)}~,  \label{4.22a}\\
\delta G &=&\frac{1}{3}D^{\a}\bar{D}^{\b}g_{\a\b}+\mri \pde^{\b(2)}g_{\b(2)}
=-\frac{1}{4}\bar{D}^{2}D^{\b}L_{\b}~,
\eea
\end{subequations}
where  $g_{\a\b} = \bar{D}_{(\a}L_{\b )}$ and 
the spinor gauge parameter $L_\a$ is an unconstrained
complex superfield.
The functional \eqref{eq:FlatAction3DLongitudinal1} coincides with 
the linearised action for (1,1) (or type I) minimal $\cN=2$ 
Poincar\'e supergravity  \cite{KT-M11}.

Varying the action \eqref{eq:FlatAction3DLongitudinal1}
with respect to the gravitational superfield $H^{\a\b}$, 
one obtains the gauge-invariant field strength
\bea
E^{\parallel}_{\a\b}&=&
\frac{1}{4}D^{\g}\bar{D}^{2}D_{\g}H_{\a\b}
-\frac{1}{8}[D_{(\a} \bar D_{\b )}][D^{\g},\bar{D}^{\d}]H_{\g\d}
-\pde_{\a\b}\pde^{\g\d} H_{\g\d}-\frac{\ri}{2}\pde_{\a\b}(G-\bar{G})~.~~~~~
\label{4.21}
\eea
Since every chiral or antichiral superfield is annihilated by 
the operator $\D$, eq. \eqref{Delta}, from \eqref{4.21} we derive 
the descendant 
\bea
W_{\a \b} (H):= -\D E^{\parallel}_{\a\b} 
=\D \Big\{ 2 \D^2 H_{\a\b} +\frac{1}{8}[D_{(\a} \bar D_{\b )}][D^{\g},\bar{D}^{\d}]H_{\g\d}
+\pde_{\a\b}\pde^{\g\d} H_{\g\d}\Big\}~, 
\label{4.22}
\eea
which is constructed solely in terms of the gravitational superfield $H_{\a\b}$. 
The gauge-invariant superfield $W_{\a\b}$ proves to be a linearised form 
of the $\cN=2$ super-Cotton tensor \cite{K12,ZupnikPak}.
The linearised expression \eqref{4.22} was recently given in \cite{CDFKS}.
Our analysis shows that the gauge-invariant field strength $W_{\a\b}$  
naturally follows from the results of the
earlier work \cite{KT-M11}.

In the supergravity framework, the super-Cotton tensor transforms homogeneously 
under the super-Weyl transformations \cite{K12} 
(see also \cite{BKNT-M} for a more general supergravity formulation).
A direct consequence of this result is that the linearised version of 
the super-Cotton tensor $W_{\a\b}$, given by eq. \eqref{4.22}, is a primary superfield 
with respect to the superconformal group.

It is an instructive exercise to show that 
\begin{align}
\Delta E^{\parallel}_{\a\b}
=&-\frac{1}{2} \D \Big\{ \Box H_{\a\b}
+ \pde_{\a}{}^{\g}\pde_{\b}{}^{\d}H_{\g\d}
+2\Delta \pde^\g{}_{(\a}H_{\b)\g} \Big\} ~.&
\end{align}
Using this relation gives an alternative expression for the field strength 
\eqref{4.22}.

Direct calculations show that $W_{\a\b}$ is  transverse linear,
\bea
\bar D^\b W_{\a\b} =  D^\b W_{\a\b} =0~.
\label{CSG4.26}
\eea
This relation is a linearised form of the Bianchi identity for the $\cN=2$ 
super-Cotton tensor \cite{BKNT-M}. It follows from \eqref{CSG4.26} that the functional
\bea
S_{\rm CSG}= \int \mrd^{3|4}z\,H^{\a\b} W_{\a\b}(H)
\label{CSG4.27}
\eea
is invariant under the gauge transformation \eqref{4.22a}.
This functional is a linearised version \cite{CDFKS} of the $\cN=2$ conformal supergravity 
action \cite{RvanN86,BKNT-M2}. 

Let us now look at  the transverse formulation for the $s=1$ case.
It is given by the following action
\bea
S^{\perp}_{{3}/{2}}[H,\Gamma, \bar \G]
&=& -\hf
\int \mrd^{3|4}z\,
\bigg\{\frac{1}{8}H^{\a\b}D^{\gamma}\bar{D}^{2}D_{\gamma}H_{\a\b}
+H^{\a\b}\left(D_{\a}\bar{D}_\b \Gamma - \bar{D}_{\a} D_{\b}\bar{\Gamma}\right)\nonumber \\
&& \phantom{-\hf
\int \mrd^{3|4}z\,\bigg\{}
+\bar{\Gamma} \Gamma 
+ \frac{3}{2}\left(\Gamma^{2}+\bar{\Gamma}^{2}\right)\bigg\}~,
\label{eq:FlatAction3DTrans1}
\eea
which is invariant under the following gauge transformations:
\begin{subequations}
\begin{align}
\delta H_{\a \b}:=& g_{\a \b}+\bar{g}_{\a \b}
=\bar{D}_{(\a}L_{\b)}-D_{(\a}\bar{L}_{\b)}~,
\label{eq:GTransHFlat1} &\\
\delta \Gamma=&\frac{1}{3}\bar{D}^{\a}D^{\b}\bar{g}_{\a\b}
=-\frac{1}{4}\bar{D}^{\b}D^{2}\bar{L}_{\b}~.
\label{eq:GTransGammaFlat1} &
\end{align}
\end{subequations}
The functional \eqref{eq:FlatAction3DTrans1}
coincides with the linearised action for $w =-1$ non-minimal 
$\cN = 2$ supergravity [13].

Associated with the action \eqref{eq:FlatAction3DTrans1} 
are the  gauge-invariant field strengths
\begin{subequations}\label{eq:HSFSHalfIntPrimaryOps1}
\begin{align}
E^{\perp}_{\a\b}
=&\frac{1}{4}D^{\gamma}\bar{D}^{2}D_{\gamma}H_{\a\b}
+D_{(\a }\bar{D}_{\b)}\Gamma - \bar{D}_{(\a }D_{\b )}\bar{\Gamma} ~,&\\
F_{\a}=&-\frac{1}{2}\bar{D}^{2}D^{\b}H_{\a\b}+\bar{D}_{\a} (\bar{\Gamma}+3\Gamma )~,&
\end{align}
\end{subequations}
in terms of which the equations of motion are $E^{\perp}_{\a\b} =0$ 
and $F_\a =0$. 
The  linearised super-Cotton tensor \eqref{4.22} 
can be expressed in terms of the field strengths \eqref{eq:HSFSHalfIntPrimaryOps1}
as follows:
\begin{align}
W_{\a\b}=&\hf \Delta E^{\perp}_{\a\b }
+\frac{\ri}{32}[D_{(\a },\bar{D}_{\b )}]\left(D^{\g}F_{\g}+\bar{D}^{\g}\bar{F}_{\g}\right)
-\frac{1}{8}\pde_{\a\b}\left(D^{\g}F_{\g}-\bar{D}^{\g}\bar{F}_{\g}\right) ~.&
\end{align}


\section{Component analysis}

The linearised gauge transformations 
\eqref{eq:GTransHFlat}--\eqref{eq:GTransGFlat} 
make use of the longitudinal linear parameter $g_{\a(2s)}$, 
given by eq. \eqref{g4.6}, and its conjugate $\bar g_{\a(2s)}$.  
The most general expression for $g_{a(2s)}$
 as  a power series in the Grassmann variables $\q$ and $\bar \q$, is 
\bea
g_{\a(2s)} (\q, \bar \q) = \re^{\ri \cH_0} \Big\{ g_{\a_1\dots \a_{2s}} 
&+& \bar \q_{(\a_1} \x_{\a_2 \dots \a_{2s})} + \q^\b \u_{\a_1 \dots \a_{2s}, \b}
+\q^2 f_{\a_1 \dots \a_{2s}} \non \\
& +& \q^\b \bar \q_{(\a_1} \l_{\a_2 \dots \a_{2s}), \b} 
+ \q^2 \bar \q_{(\a_1} \S_{\a_2 \dots \a_{2s})} \Big\}~,
\label{com5.1}
\eea
where 
\bea 
\cH_0 := \q^\a (\g^m)_{\a\b} \bar \q^\b \pa_m = \q^\a\bar \q^\b \pa_{\a\b} \equiv 
\r^{\a\b} \pa_{\a\b} ~, \qquad \r^{\a\b} := \q^{(\a} \bar \q^{\b)}~.
\label{com5.2}
\eea
All component fields in \eqref{com5.1} are complex and symmetric in their 
$\a$-indices. 
The components $\U_{\a_1 \dots \a_{2s}, \b}$ 
and $ \l_{\a_1 \dots \a_{2s-1}, \b} $ are not required to have 
any symmetry property relating their $\a$ and 
 $\b$ indices, which is indicated by a coma. In other words,
$\U_{\a (2s), \b}$ belongs to the tensor product 
$(\bf{2s+1}) \otimes\bf 2$ of two $\sSL (2,{\mathbb R})$ representations.

As follows from the  gauge transformation  \eqref{eq:GTransHFlat}, 
the component gauge parameters $g_{\a(2s)}$, $\U_{\a(2s), \b} $ 
and $f_{\a(2s)}$  in \eqref{com5.1} can be used to choose a Wess-Zumino gauge
of the form:
\bea
H_{\a_1 \dots \a_{2s}} (\q,\bar \q)&=& 
\ri \,\q\bar \q \,D_{\a_1 \dots \a_{2s}} 
+ \q^{(\b} \bar \q^{\g)} E_{ \a_1 \dots \a_{2s}, \b\g}
+\bar \q^2 \q^\b \J_{ \a_1 \dots \a_{2s}, \b}
-\q^2 \bar \q^\b \bar \J_{\a_1 \dots \a_{2s}, \b} \non \\
&& + \q^2 \bar \q^2 A_{\a_1 \dots \a_{2s} }~, 
\label{com5.3}
\eea
where the composite scalar $\q\bar \q = \q^\a \bar \q_\a$ is imaginary. 
 All bosonic fields in \eqref{com5.3} are real. 
 So far  no gauge condition has been imposed on 
 $\G_{\a_1 \dots \a_{2s-2}} $.
 To preserve the gauge condition \eqref{com5.3}, 
 some of the gauge parameters contained in \eqref{com5.1} must be  
 constrained as follows:
 \begin{subequations}
 \bea
g_{\a_1 \dots \a_{2s} } &=& -\frac{\ri}{2}  \z_{\a_1 \dots \a_{2s} } ~, \qquad 
\bar \z_{\a(2s)} = \z_{\a(2s)} ~, \label{com5.4a}
\\
f_{\a_1 \dots \a_{2s} } &=& 0~, \\
\u_{\a_1 \dots \a_{2s}, \b} &=& - \ve_{\b(\a_1} \bar \x_{\a_2 \dots \a_{2s})}~.
 \eea
 \end{subequations}

The first term in the second line of \eqref{com5.1} can be represented as 
\bea
 \q^\b \bar \q_{(\a_1} \l_{\a_2 \dots \a_{2s}), \b}  
 = \hf \q \bar \q \L_{\a_1 \dots \a_{2s}}
 +\frac{2s+1}{2s} \r^\b{}_{(\a_1}\L_{\a_2 \dots \a_{2s} \b)} 
 -\frac{2s-1}{2s} \r_{(\a_1 \a_2} \L_{\a_3 \dots \a_{2s})} ~, ~~~
 \label{com5.5}
 \eea
 where we have introduced two irreducible components of 
 $\l_{\a_1 \dots \a_{2s-1}, \b} $ by the rule
 \bea
 \L_{\a_1 \dots \a_{2s}} := \l_{(\a_1 \dots \a_{2s-1}, \a_{2s})}  ~, 
 \qquad
 \L_{\a_2 \dots \a_{2s-2}}  := \l_{\a_2 \dots \a_{2s-2}\b,}{}^\b  ~.
 \eea
We recall that the composite $\r^{\a\b}$ is defined by \eqref{com5.2}.
It is clear from  \eqref{eq:GTransHFlat}, \eqref{com5.1}, 
\eqref{com5.3} and \eqref{com5.5} that the imaginary part of $ \L_{\a(2s)} $
can be used to gauge away the component field $D_{\a(2s)}$ thus arriving 
at the stronger Wess-Zumino gauge
\bea
H_{\a_1 \dots \a_{2s}} (\q,\bar \q)= 
 \q^{(\b} \bar \q^{\g)} E_{ \a_1 \dots \a_{2s}, \b\g}
&+&\bar \q^2 \q^\b \J_{ \a_1 \dots \a_{2s}, \b}
-\q^2 \bar \q^\b \bar \J_{\a_1 \dots \a_{2s}, \b} \non \\
&+& \q^2 \bar \q^2 A_{\a_1 \dots \a_{2s} }~, 
\label{com5.7}
\eea
in which the residual $\L$-invariance is described by a real parameter,
\bea
\bar \L_{\a(2s) } = \L_{\a(2s)}\equiv l_{\a(2s)} ~.
\label{com5.8}
\eea 

The real bosonic field $E_{ \a_1 \dots \a_{2s}, \b\g}$ 
transforms in the representation $(\bf{2s+1}) \otimes\bf 3$ of $\sSL (2,{\mathbb R})$, 
while the complex fermionic field $\J_{ \a_1 \dots \a_{2s}, \b}$ belongs to 
the $(\bf{2s+1}) \otimes\bf 2$. The field $E_{ \a(2s), \b\g}$ is a higher spin 
analogue of the linearised vielbein (or frame field), which becomes obvious
if we convert the spinor indies  of  $E_{\a(2s), \b\g}$ into vector ones
by the standard rule 
\bea
E_m{}^{a_1 \dots a_s} := \Big( -\hf \Big)^{s+1} 
(\g^{a_1})^{\a_1\a_2} \dots (\g^{a_s})^{\a_{2s-1}\a_{2s}}
(\g_m)^{\b\g}E_{ \a_1 \dots \a_{2s}, \b\g}~.
\eea 
Here $E_m{}^{a_1 \dots a_s}$ is symmetric and traceless with respect to 
the  indices $a_1,  \dots , a_s$.
This interpretation is confirmed by the fact that the gauge transformation associated with 
the parameter  \eqref{com5.4a} acts on 
$E_{ \a_1 \dots \a_{2s}, \b\g}$ as follows:
\bea
\d E_{ \a_1 \dots \a_{2s}, \b\g} = \pa_{\b\g} \z_{\a_1\dots \a_{2s}} \quad \Longleftrightarrow
\quad \d E_m{}^{a_1 \dots a_s} = \pa_m \z^{a_1 \dots a_s}~.
\label{com5.10}
\eea
The gauge transformation generated by the parameter \eqref{com5.8}
acts on $E_{ \a_1 \dots \a_{2s}, \b\g} $ as a higher spin counterpart of the linearised
local Lorentz transformation. Therefore the tensor structure of 
$E_m{}^{ a \dots \a_{s}}$ and its gauge freedom correspond to the 3D massless spin-$(s+1) $
gauge field, see e.g. \cite{Vasiliev}. In the frame-like formulation for massless higher spin 
fields \cite{Vasiliev}, one introduces two independent gauge fields, one of which is 
$E_m{}^{ a \dots \a_{s}}$ and the other is  a higher spin analogue of the Lorentz connection.
The latter is expressed in terms of $E_m{}^{ a \dots \a_{s}}$ on the equations of motion. 
However, in the off-shell formulations for supergravity, 
no independent Lorentz connection appears. 
And its higher spin analogue never appears in the framework 
of off-shell higher spin supermultiplets \cite{KSP,KS93,KS94}.

Recalling the gauge transformation law of $\G_{\a(2s-2)}$, eq. \eqref{eq:GTransGammaFlat}, 
one may see that  the gauge freedom associated with the parameters $\L_{\a(2s-2) }$  
in \eqref{com5.5} and  $\S_{\a(2s-1)}$ in \eqref{com5.1} allows us to bring $\G_{\a(2s-2)}$
to the following form:
\bea
\G_{\a_1 \dots \a_{2s-2}} (\q, \bar \q) &=& \re^{\ri \q^\l {\bar \q}{}^\r \pa_{\l\r} } \Big[
\q^\b \o_{(\b \a_1 \dots \a_{2s-2})} 
+\q_{(\a_1 } \bar \J_{\a_2 \dots \a_{2s-2})} 
\non \\
&&{}\qquad \quad 
+\q^2 B_{\a_1 \dots \a_{2s-2} } + \bar \q^\b  \q^\g U_{(\b\g \a_1 \dots \a_{2s-2})} 
+\bar \q^\b \q_{ (\b} F_{\a_1 \dots \a_{2s-2} )}
\non \\
&&{}\qquad \quad 
+ \q^2 \bar \q^\b \r_{(\b \a_1 \dots \a_{2s-2})} \Big]~.
\label{com5.11}
\eea
The fermionic fields $ \J_{ \a_1 \dots \a_{2s}, \b}$ in \eqref{com5.7} 
and $\J_{\a_1 \dots \a_{2s-3} }$ appearing in $\bar \G_{\a_1 \dots \a_{2s-2}} $
constitute a complex version of the massless spin-$(s+\hf)$ field reviewed in Appendix 
\ref{AppendixB.2}. The complex fermionic fields $\o_{\a(2s-1)} $ and $\r_{\a(2s-1)} $
in \eqref{com5.11} turn out to be auxiliary for the theory with action \eqref{eq:FlatAction3DTrans}
in the standard sense that they become 
functions of the other fermionic fields on the mass shell. 
The real bosonic field $A_{\a (2s) }$  in \eqref{com5.7} 
and the complex bosonic fields $B_{\a(2s-2) } $, $U_{\a (2s) }$ and $F_{\a (2s-2)}$ 
in \eqref{com5.11}  are auxiliary for the theory with action \eqref{eq:FlatAction3DTrans}. 

Now we are can argue that, upon elimination of the auxiliary fields, 
the  theory with action \eqref{eq:FlatAction3DTrans} is 
equivalent to a sum of two massless (Fang-)Fronsdal models, one of which  is the bosonic
spin-$(s+1)$ model described in Appendix B.1 and the other corresponds to 
two identical fermionic spin-$(s+1/2)$ models described in Appendix B.2. 
Equivalently, the fermionic sector describe a complex massless spin-$(s+1/2)$ gauge field. 
Indeed, consider the frame field in \eqref{com5.7}. It can be represented as the sum 
of three irreducible components, 
\bea
 \r^{\b \g} E_{ \a_1 \dots \a_{2s}, \b\g} = \r^{\b\g}  h_{ \a_1 \dots \a_{2s} \b\g}
 +\r^\b{}_{(\a_1} m_{\a_2 \dots \a_{2s} ) \b} + \r_{(\a_1 \a_2} h_{\a_3\dots \a_{2s})}~,
\eea
where the irreducible components of $E_{ \a_1 \dots \a_{2s}, \b\g} $
are defined by 
\begin{subequations}
\bea
h_{\a_1 \dots \a_{2s +2} }&:=& E_{ (\a_1 \dots \a_{2s}, \a_{2s+1} \a_{2s+2}) }~, \\
m_{\a_1 \dots \a_{2s}} &:=& -\frac{2s+1}{s+1} E_{ (\a_1 \dots \a_{2s}, \b) }{}^\b ~, \\
h_{\a_1 \dots \a_{2s-2} }&:=& \frac{2s-1}{2s+1} E_{\a_1 \dots \a_{2s-2} \b\g,}{}^{\b\g} ~.
\eea
\end{subequations}
The field $m_{\a(2s)} $ may be algebraically gauged away by the generalised 
Lorentz transformation described by the parameter \eqref{com5.8}. 
The remaining bosonic fields $h_{\a(2s+2)}$ and $h_{\a(2s-2)}$ correspond to the dynamical variables of the Fronsdal spin-$(s+1)$ model reviewed in Appendix B.1. 
As follows from \eqref{com5.10}, their gauge freedom is equivalent to that of the massless
spin-$(s+1)$ gauge field, see eqs. \eqref{B.2}. Since the requirement of gauge invariance fixes
the Fronsdal action modulo an overall numerical factor, we are confident the  theory \eqref{eq:FlatAction3DTrans} leads to the Fronsdal spin-$(s+1)$ model even without 
explicit calculation of the component bosonic action. Such a calculation may be carried out 
in complete analogy with the 4D case described in \cite{KS94}; we will not consider it here. 
 Let us turn to the fermionic sector and represent the higher-spin gravitino 
$\J_{ \a_1 \dots \a_{2s}, \b} $ in \eqref{com5.7} as the sum 
of two irreducible components, 
\bea
\J_{ \a_1 \dots \a_{2s}, \b} = \J_{ \a_1 \dots \a_{2s} \b } + \ve_{\b (\a_1} \J_{\a_2 \dots \a_{2s})} ~,
\eea
where the irreducible components of 
are defined by 
\begin{subequations}
\bea
\J_{ \a_1 \dots \a_{2s} \b } &:=& \J_{ (\a_1 \dots \a_{2s} ,\b )}~, \\
\J_{ \a_1 \dots \a_{2s-1} } &:=& \frac{2s}{2s+1} \J_{ \a_1 \dots \a_{2s-1} \b, }{}^\b~.
\eea
\end{subequations}
These complex fermionic fields $\J_{\a(2s+1)}$, $\J_{\a(2s-1)}$ 
as well as  $\J_{\a(2s-3) }$ sitting in $\bar \G_{ \a (2s-2)} $
constitute a complex version of the massless spin-$(s+\hf)$ field reviewed in Appendix 
\ref{AppendixB.2}. Under the fermionic local symmetry generated by the 
complex parameter $\x_{\a(2s-1)}$ in \eqref{com5.1}, 
the  gauge transformation law of these fields is equivalent to the complex  version 
of the transformation \eqref{B.13} which corresponds to the massless spin-$(s+1/2)$
field. Since the requirement of gauge invariance fixes
the Fang-Fronsdal action modulo an overall numerical factor, we are confident the  theory \eqref{eq:FlatAction3DTrans} leads to the massless spin-$(s+1/2)$ model even without 
explicit calculation of the component fermionic action.

Instead of dealing with the gauge \eqref{com5.7} and \eqref{com5.11}, 
sometimes it is more convenient to work with an alternative Wess-Zumino gauge
defined by 
\begin{subequations} \label{eq:WZGauge_new}
\bea
H_{\a_1 \dots \a_{2s}} (\q,\bar \q)&=& \q^\b \bar \q^\g h_{(\b \g \a_1 \dots \a_{2s})}
+\bar \q^2 \q^\b \J_{(\b \a_1 \dots \a_{2s})}
-\q^2 \bar \q^\b \bar \J_{(\b \a_1 \dots \a_{2s})} \non \\
&& + \q^2 \bar \q^2 A_{\a_1 \dots \a_{2s} }~, \\
\G_{\a_1 \dots \a_{2s-2}} (\q, \bar \q) &=& \re^{\ri \q^\l {\bar \q}{}^\r \pa_{\l\r} } \Big[
h_{\a_1 \dots \a_{2s-2}} + \q^\b \J_{(\b \a_1 \dots \a_{2s-2})} 
+\q_{(\a_1 } \bar \J_{\a_2 \dots \a_{2s-2})} \non \\
&&{}\qquad \quad + \bar \q^\b \U_{(\b \a_1 \dots \a_{2s-2})}
+\q^2 B_{\a_1 \dots \a_{2s-2} } + \bar \q^\b  \q^\g U_{(\b\g \a_1 \dots \a_{2s-2})} \non \\
&&{}\qquad \quad +\bar \q^\b \q_{ (\b} F_{\a_1 \dots \a_{2s-2} )}
+ 2\q^2 \bar \q^\b {\bm \r}_{(\b \a_1 \dots \a_{2s-2})}
\Big]~,
\eea
with
\bea
{\bm \rho}_{\a_1 \dots \a_{2s-1}}=\rho_{\a_1 \dots \a_{2s-1}}
-\frac{\ri}{4} \pde^\b{}_{(\a_1} \psi_{\a_2 \dots \a_{2s-1})\b}
-\frac{\ri}{4}\pde_{(\a_1\a_2}\psi_{\a_2 \dots \a_{2s-1})}~.
\eea
\end{subequations}
Here the bosonic fields $h_{\a(2s) }$, $h_{\a(2s-2) }$ and $A_{\a (2s) }$ are real, 
and the fields $B_{\a(2s-2) } $, $U_{\a (2s) }$ and $F_{\a (2s-2)}$ are complex.


\section{Superconformal higher spin multiplets}

In this section we develop a superspace setting for 
linearised higher spin conformal supergravity. 
We  start with a review of a review of the conformal Killing supervector fields
of 3D $\cN=2$ Minkowski superspace  \cite{Park3,KPT-MvU}, which are defined in complete
analogy with the 4D $\cN=1$ case \cite{BK}.

\subsection{Conformal Killing supervector fields} 

Consider a real supervector field $\x$ on Minkowski superspace, 
\bea
\x=   \x^B D_B := \x^b \pa_b + \x^\b D_\b +\bar\x_\b \bar D^\b 
= -\hf \x^{\b\g} \pa_{\b\g} + \x^\b D_\b +\bar\x_\b \bar D^\b ~.
\eea
It is called a conformal Killing supervector field if it 
obeys the equation 
\bea
[\x +K^{\b\g}M_{\b\g} , D_\a] + \d_\r D_\a =0 \quad \Longleftrightarrow \quad
[\x +K^{\b\g}M_{\b\g} , \bar D_\a] + \d_\r \bar D_\a =0~,
\label{5.2}
\eea
for some Lorentz ($K^{\b\g} = K^{\g\b} = \bar K^{\b\g}$) and super-Weyl 
($\r$) parameters. 
We recall that the Lorentz generator $M_{\b\g}$  acts on a spinor $\j_\a$ by the rule
\bea
M_{\b\g} \j_\a = \ve_{\a (\b} \j_{\g)} = \hf ( \ve_{\a \b} \j_\g + \ve_{\a\g} \j_\b)~.
\eea
The super-Weyl transformation  of the covariant derivatives is defined according to 
\cite{KT-M11}
\begin{subequations}
\bea
\d_\r D_\a &=& \hf(3\bar \r -\r) D_\a + (D^\l \r) M_{\l \a}~, \\
\d_\r \bar D_\a &=& \hf(3 \r - \bar \r) \bar D_\a + (\bar D^\l \bar \r) M_{\l \a}~,
\eea
where the parameter $\r$ is chiral,
\bea
\bar D_\a \r=0~.
\eea
\end{subequations}

Eq. \eqref{5.2} can be rewritten in the form \cite{KPT-MvU}
\bea
[\x , D_\a ] =- K_\a{}^\b D_\b +\hf (\r - 3\bar \r) D_\a~.
\label{5.5}
\eea
The equation \eqref{5.2}, or its equivalent form \eqref{5.5}, implies
\bea
D_\a \x^{\b \g} + 4\ri \d_\a^{(\b} \bar \x^{\g)} =0  \quad \Longleftrightarrow \quad
\bar D_\a \x^{\b \g} - 4\ri \d_\a^{(\b}  \x^{\g)} =0~,
\eea
and therefore the spinor components $\x^\a$ and $\bar \x_\a$ of $\x$ 
are determined in terms of the vector ones,
\bea
\x^\a = -\frac{\ri}{6} \bar D_\b \x^{\a\b}~, 
\qquad \bar \x_\a = -\frac{\ri}{6}  D^\b \x_{\a\b}~,
\eea
and the vector component $\x^{\a\b} = \x^{\b \a} = \bar \x^{\a\b}$ is longitudinal linear,
\bea
D^{(\a}\x^{\b\g)}=0~, \qquad \bar D^{(\a}\x^{\b\g)}=0~,
\eea
and therefore $\x^{\a\b}$ is an ordinary conformal Killing vector, 
\bea
\pa^{(\a\b} \x^{\g\d)} =0~.
\eea
These relations imply that $D^2 \x^{\a\b} = \bar D^2 \x^{\a\b}=0$, and therefore
$\x^\a$ is chiral, 
\bea
\bar D_\a \x^\b =0~.
\label{5.10}
\eea

It follows from  \eqref{5.2}, or its equivalent form \eqref{5.5}, 
that the Lorentz and super-Weyl parameters, $K_{\a\b}$ and $\r$, 
are uniquely expressed in terms of the components of the conformal Killing 
supervector field as follows:
\begin{subequations}
\bea
K_{\a\b}&=& D_{(\a}\x_{\b)} = - \bar D_{(\a} \bar \x_{\b)}~, \\
\r&=& \frac{1}{8}(D_\a \x^\a +3 \bar D^\a \bar \x_\a)~.
\label{5.11b}
\eea
\end{subequations}
We also deduce from \eqref{5.2} that the Lorentz and super-Weyl 
parameters are related to each other as
\bea
D_\a K^{\b\g} = \d_\a{}^{(\b } D^{\g)}\r~.
\eea
Using the properties \eqref{5.5} -- \eqref{5.10}, 
one can explicitly check that $\r$ defined by \eqref{5.11b} is chiral. 

\subsection{Primary linear superfields}

A symmetric rank-$n$ spinor superfield $\F_{\a_1 \dots \a_n} $ 
is said to be primary of dimension $\hf(x+y)$ 
if its superconformal transformation is 
\bea
\d_\x \F_{\a_1 \dots \a_n} = \x \F_{\a_1 \dots \a_n} 
+n K^\b{}_{(\a_1} \F_{\a_2 \dots \a_n)\b } + (x\r +y \bar \r) \F_{\a_1 \dots \a_n} ~.
\label{5.13}
\eea
for some real parameters $x$ and $y$. 
The $R$-charge of $\F_{\a(n)} $ is proportional to  $\hf(x-y)$. 

Let $G_{\a_1 \dots \a_n} $ be a longitudinal linear superfield constrained by 
\eqref{LL}.  Requiring $G_{\a_1 \dots \a_n} $ to be primary fixes 
one of the superconformal parameters in \eqref{5.13}, 
\bea
\d_\x G_{\a_1 \dots \a_n} = \x G_{\a_1 \dots \a_n} 
+n K^\b{}_{(\a_1} G_{\a_2 \dots \a_n)\b } 
+ \left(x\r -\frac{n}{2} \bar \r \right) G_{\a_1 \dots \a_n} ~.
\label{LLprimary}
\eea
Let $\G_{\a_1 \dots \a_n} $ be a transverse linear superfield constrained by 
\eqref{TL}.  Requiring $\G_{\a_1 \dots \a_n} $ to be primary fixes 
one of the superconformal parameters in \eqref{5.13}, 
\bea
\d_\x \G_{\a_1 \dots \a_n} = \x \G_{\a_1 \dots \a_n} 
+n K^\b{}_{(\a_1} \G_{\a_2 \dots \a_n)\b } 
+ \left( x\r + (1+\frac{n}{2} )\bar \r \right) \G_{\a_1 \dots \a_n} ~.
\label{TLprimary}
\eea
An analysis of constrained primary superfields were given by Park \cite{Park3}. 

Now, let us come back to the gauge transformation law \eqref{eq:GTransHFlat}. 
We postulate that the real gauge prepotential $H_{\a(2s)} $ and the right-hand  side of \eqref{eq:GTransHFlat} are primary.  Then it follows from \eqref{LLprimary} 
that the superconformal transformation of $H_{\a(2s)} $ is 
\bea
\d H_{\a_1 \dots \a_{2s}} = \x H_{\a_1 \dots \a_{2s}} 
+2s K^\b{}_{(\a_1} H_{\a_2 \dots \a_{2s})\b } 
-  s \left(\r + \bar \r \right) H_{\a_1 \dots \a_{2s}} ~,
\label{Hprimary}
\eea
and the dimension of $H_{\a(2s)}$ is equal to $(-s)$. 

Let $W_{\a_1 \dots \a_n}$ be a real transverse linear superfield, 
\bea
D^\b W_{\b \a_1 \dots \a_{n-1} } = \bar D^\b W_{\b \a_1 \dots \a_{n-1} } = 0~.
\label{6.177}
\eea
Requiring it to be primary, we deduce from \eqref{TLprimary} that 
the superconformal transformation of $W_{\a(n)} $ is 
\bea
\d W_{\a_1 \dots \a_{n}} = \x W_{\a_1 \dots \a_{n}} 
+n K^\b{}_{(\a_1} W_{\a_2 \dots \a_{n})\b } 
+\left(1 + \frac{n}{2}\right) (\r + \bar \r ) W_{\a_1 \dots \a_{n}} ~,
\label{Wprimary}
\eea
and the dimension of $W_{\a(n)}$ is equal to $(1+n/2)$. 

\subsection{Linearised higher spin conformal supergravity}

Consider an action of the form 
\bea
S = \ri^n \int \mrd^{3|4}z\, H^{\a_1 \dots \a_{n}}W_{\a_1 \dots \a_{n}}~,
\label{5.18}
\eea
where $H_{\a(n)}$ is a real symmetric rank-$n$ spinor superfield
with the superconformal transformation 
\bea
\d H_{\a_1 \dots \a_{n}} = \x H_{\a_1 \dots \a_{n}} 
+ K^\b{}_{(\a_1} H_{\a_2 \dots \a_{n})\b } 
-  \frac{n}{2} \left(\r + \bar \r \right) H_{\a_1 \dots \a_{n}} ~,
\label{Hprimary2}
\eea
which coincides with \eqref{Hprimary} for $n=2s$.
The action \eqref{5.18} is invariant under the superconformal transformations
\eqref{Wprimary} and \eqref{Hprimary2}. 
Moreover, it is also invariant under gauge transformations of the form 
\bea
\d H_{\a(n)} = g_{\a(n)} + \bar g_{\a(n)}~, \qquad \d W_{\a_1\dots  \a_n}=0~, \qquad
g_{\a_1\dots \a_n} = \bar{D}_{(\a_{1}}L_{\a_{2}...\a_{n})} ~,
\label{5.19}
\eea
where the complex gauge parameter  $g_{\a(n)}  $ 
is an arbitrary longitudinal linear superfield. 
The gauge invariance follows from \eqref{6.177}.
This  gauge transformation law reduces to \eqref{eq:GTransHFlat}
for $n =2s$. We would like to realise $W_{\a(n)}$ as a gauge-invariant field strength,
$W_{\a(n)} (H)$, constructed from the prepotential $H_{\a(n)}$.
Then \eqref{5.18} may be interpreted as a higher spin extension of the linearised conformal
supergravity action \eqref{CSG4.27}.

Given a prepotential $H_{\a(n)} = H_{\a_1 \dots \a_n}$ 
with the superconformal transformation law \eqref{Hprimary}
and the gauge transformation 
\eqref{5.19}, we associate with it a  gauge-invariant real field strength 
$W_{\a(n)} (H) $ defined by 
\bea
&&W_{\a_1 \dots \a_n}  (H)
:= \frac{1}{2^{n-1}} 
\sum\limits_{J=0}^{\left \lfloor{n/2}\right \rfloor}
\bigg\{
\binom{n}{2J} 
\Delta  \Box^{J}\pde_{(\a_{1}}{}^{\b_{1}}
\dots
\pde_{\a_{n-2J}}{}^{\b_{n-2J}}H_{\a_{n-2J+1}\dots\a_{n})\b_1 \dots\b_{n-2J}}~~~~
\nonumber \\
&&\qquad \qquad +
\binom{n}{2J+1}\Delta^{2}\Box^{J}\pde_{(\a_{1}}{}^{\b_{1}}
\dots\pde_{\a_{n-2J -1}}{}^{\b_{n-2J -1}}H_{\a_{n-2J}\dots\a_{n})
\b_1 \dots \b_{n-2J -1} }\bigg\}~,~~~~~
\label{eq:HSFSUniversal}
\eea
where $\left \lfloor{x}\right \rfloor$ denotes the floor (also known as the integer part) of 
a number $x$. 
One may check that 
\bea
 \int \mrd^{3|4}z\, \widetilde{H}^{\a_1 \dots \a_{n}}W_{\a_1 \dots \a_{n}} (H)
 =\int \mrd^{3|4}z\, {H}^{\a_1 \dots \a_{n}}W_{\a_1 \dots \a_{n}} (\widetilde{H})~,
 \eea
for arbitrary superfields $H_{\a(n)} $ and $\widetilde{H}_{\a(n)}$ that are 
 bosonic for even $n$ and fermionic for odd $n$. 
 The field strength obeys the Bianchi identities
 \bea
D^{\b}W_{\b\a_1 \dots \a_{n-1}}=0, \qquad 
\bar{D}^{\b}W_{\b\a_1 \dots \a_{n-1}}=0 ~.
\label{6.244}
\eea
These Bianchi identities are compatible only 
with the superconformal transformation law \eqref{Wprimary}, 
and thus $W_{\a(n)}$ is a primary superfield. 
In Appendix \ref{AppendixD}, we prove (i)  invariance of 
the field strength \eqref{eq:HSFSUniversal} under the gauge transformation 
\eqref{5.19}; and (ii) the Bianchi identies \eqref{6.244}.

In the case of the half-integer superspin 
transverse formulation, the superconformal field strength 
$W_{\a(2s)}$  can be expressed in terms of the gauge-invariant field strengths
\eqref{4.11ab} as follows:
\bea
2^{2s} W_{\a(2s)}
&=&\phantom{-} \frac{1}{2}\pde_{(\a_{1}}{}^{\b_{1}}....\pde_{\a_{2s-2}}{}^{\b_{2s-2}}\pde_{\a_{2s-1}\a_{2s})}
\Big(D^{\gamma}F_{\gamma\b(2s-2)} -\bar{D}^{\gamma}\bar{F}_{\gamma\a(2s-2)}\Big) \non \\
&&-\frac{1}{2}\binom{2s}{2}\pde_{(\a_{1}}{}^{\b_{1}}....\pde_{\a_{2s-2}}{}^{\b_{2s-2}}E^{\perp}_{\a_{2s-1}\a_{2s})\b(2s-2)}\nonumber \\
&&-\sum\limits_{J=1}^{s}\binom{2s}{2J}\Delta\Box^{J-1}\pde_{(\a_{1}}{}^{\b_{1}}....\pde_{\a_{2s-2J}}{}^{\b_{2s-2J}}
E^{\perp}_{\a_{2s-2J+1}...\a_{2s})\b(2s-2)}
\nonumber \\
&&-2\sum\limits_{J=0}^{s-1}\binom{2s}{2J+1}\pde_{(\a_{1}}{}^{\b_{1}}\dots\pde_{\a_{2s-2J-1}}{}^{\b_{2s-2J-1}}\Box^{J}E^{\perp}_{\a_{2s-2J}...\a_{2s})\b(2s-2J-1)}\label{eq:HSFSHalfInteger2}
~.~~~~~
\eea
This gauge-invariant field strength is a higher spin extension 
of the linearised super-Cotton tensor \eqref{4.22}.


\section{Massive half-integer superspin models}

We now consider a gauge-invariant deformation of the transverse action 
\eqref{eq:FlatAction3DTrans}
\begin{equation}
S^\perp = \m^{2s-1} S^{\perp}_{s+\frac{1}{2}} [H, \G, \bar \G] 
+\frac{\lambda}{2} S_{\rm CS} [H]~,
\label{eq:HSMassiveAction}
\end{equation}
where $S_{\rm CS} [H]$ denotes 
the Chern-Simons-type superconformal term 
\begin{equation}
S_{\rm CS} [H]=  \left(-\frac{1}{2}\right)^{s}\int \rd^{3|4}z \, H^{\a(2s)}W_{\a(2s)} (H)~,
\label{eq:HSChernSimmonsAction}
\end{equation}
with the field strength $W_{\a_(2s)} (H)$ given by \eqref{eq:HSFSUniversal}.
The coupling constant $\l$ in \eqref{eq:HSMassiveAction} is dimensionless.
In accordance with \eqref{Hprimary}, 
the dimension of $H_{\a(2s)}$ is equal to $(-s)$. To make the action dimensionless, 
the first term in \eqref{eq:HSMassiveAction} is rescaled  
by an overall factor $\m^{2s-1} $
with the positive parameter $\m$  of unit mass dimension. 

In the $s=1$ case, the action \eqref{eq:HSMassiveAction} 
coincides with the linearised action for topologically massive $\cN=2$ supergravity 
in the non-minimal $w=-1$ formulation \cite{KLRST-M}.

Since  $S_{\rm CS} [H]$
does not involve the compensating superfield $\Gamma_{\a(2s-2)}$ and its conjugate,  
it follows that the corresponding equations of motion are the same 
for both the massive \eqref{eq:HSMassiveAction}  and massless theories \eqref{eq:FlatAction3DTrans}:
\begin{equation}
F_{\a(2s-1)}=0 \quad \Longrightarrow \quad 
D^{\b}E^\perp_{\b\a(2s-1)}=0~.
\label{7.3}
\end{equation}
However, the addition of the Chern-Simons term results in the following modification 
to the $H_{\a(2s)}$ equation of motion for the massive theory:
\begin{equation}
\m^{2s-1}  E^\perp_{\a(2s)}+\lambda W_{\a(2s)}=0 ~.
\label{eq:HSFSMassiveEOM}
\end{equation}
This is a gauge-invariant higher-derivative superfield equation.\footnote{As demonstrated
in \cite{BHT}, the general second-order 3D massive field equations for positive integer spin, 
and their ``self-dual'' limit to first-oder equations, are equivalent to gauge-invariant 
higher-derivative equations.}

Once the equations of motion \eqref{7.3} hold, 
one can obtain a simplified expression for $W_{\a(2s)}$. It  is 
\begin{align}
W_{\a(2s)}=& -\Delta \Box^{s-1}E^\perp_{\a(2s)}~.&
\end{align}
Using this result, we can extract a `higher-superspin' analogue of the Klein-Gordon equation 
from the equation of motion \eqref{eq:HSFSMassiveEOM} as follows. First note that
\bea
0&=&m^{2s-1}E^\perp_{\a(2s)}+\lambda W_{\a(2s)}=
\m^{2s-1}E^\perp_{\a(2s)}- \l
\Delta \Box^{s-1}E^\perp_{\a(2s)} \quad \implies \nonumber \\
0&=&\m^{2s-1}\Delta E^\perp_{\a(2s)}-2\l \Box^{s-1}\Delta^{2}E^\perp_{\a(2s)}
= (\m^{2s-1} \Delta 
 -2 \l \Box^{s})E^\perp_{\a(2s)}~.
\eea
This leads to 
\begin{equation}
\big(\Box^{2s-1}- (m^2)^{2s-1} 
\big) E^\perp_{\a(2s)}=0~, \qquad  m := \frac{\m}{| \l|^{1/(2s-1)}}~.
\label{eq:HSMassiveKleinGordon}
\end{equation}
By making use of the Fourier transform of $E^\perp_{\a(2s)}$, we deduce from \eqref{eq:HSMassiveKleinGordon}
the ordinary Klein-Gordon equation 
\bea
\big(\Box- m^2
\big) E^\perp_{\a(2s)}=0~.
\eea
It then follows from  \eqref{eq:HSFSMassiveEOM} that 
\begin{subequations}
\bea
\D E^\perp_{\a (2s)} &=& m \s E^\perp_{\a(2s)}~, \qquad \s =\frac{\l}{|\l |}~.
\label{massiveE.a}
\eea
It remains to recall that 
\bea
D^{\b}E^\perp_{\b\a(2s-1)}= \bar D^{\b}E^\perp_{\b\a(2s-1)}=0~.
\label{massiveE.b}
\eea
\end{subequations}
The equations \eqref{massiveE.a} and \eqref{massiveE.b} tell us that 
$E^\perp_{\a(2s)}$ is a massive superfield of superhelicity 
$\k = (s+\hf)\s$, in accordance with the analysis given in section \ref{subsection3.2}.

The massive theory \eqref{eq:HSMassiveAction}
possesses a dual formulation. It is described by the action 
\begin{equation}
S^{\parallel} = \m^{2s-1} S^{\parallel}_{s+\frac{1}{2}} [H, G, \bar G] 
+\frac{\lambda}{2} S_{\rm CS} [H]~,
\label{7.8}
\end{equation}
with the longitudinal action  $S^{\parallel}_{s+\frac{1}{2}} [H, G, \bar G] $
given by eq. \eqref{eq:FlatAction3DLongitudinal}.
Since  $S_{\rm CS} [H]$
does not involve the compensating superfield $G_{\a(2s-2)}$ and its conjugate,  
the equation of motion for $G_{\a(2s-2)}$ 
is the same as in massless theory, eq. \eqref{lon4.20}. Due to the Bianchi identity 
\eqref{lon4.19}, we obtain 
\begin{equation}
B_{\a(2s-3)}=0 \quad \Longrightarrow \quad 
D^{\b}E^{\parallel}_{\b\a(2s-1)}=0~.
\end{equation}
However, the equation of motion for $H_{\a(2s)}$ becomes
\begin{equation}
\m^{2s-1}  E^\parallel_{\a(2s)}+\lambda W_{\a(2s)}=0 ~.
\end{equation}
This demonstrates that the massive models  \eqref{eq:HSMassiveAction}
and \eqref{7.8} possess equivalent dynamics. 

In the $s=1$ case, the action \eqref{7.8} 
coincides with the linearised action for topologically massive $\cN=2$ supergravity 
in the minimal (1,1) formulation \cite{KLRST-M}.

\section{Concluding comments}

The main results of this paper are as follows. In section 4 we  constructed 
the two dually equivalent off-shell formulations for the massless superspin-$(s+1/2)$ multiplet, 
with $s>1$, as 3D analogues of the  off-shell 
4D $\cN=1$ massless  multiplets of half-integer superspin \cite{KSP}. 
In section 6 we presented the linearised higher spin 
super-Cotton tensors and the linearised actions for higher spin conformal supergravity.  
In section 7 we constructed the off-shell formulation for massive
superspin-$(s+1/2)$ multiplets, with $s>1$, as higher-spin extensions 
of the off-shell  topologically massive $\cN=2$ supergravity theories \cite{KLRST-M}.  

This paper does not include any 3D analogues of the  off-shell 
4D $\cN=1$ massless  multiplets of integer superspin \cite{KS93}. 
We have constructed such extensions. However, they do not admit 
massive deformation of the type described in section 7. That is why we will 
discuss these models elsewhere. 
This paper does not include any 3D analogues 
of the off-shell 4D $\cN=1$ higher spin supermultiplets 
in anti-de Sitter space \cite{KS94}. We have constructed such extensions. 
They will be discussed elsewhere. 

To the best of our knowledge, no off-shell 3D $\cN=1$ massive higher spin supermultiplet have appeared in the literature. 
They can be derived by carrying out the plain superspace reduction $\cN=2 \to \cN=1$
to the models presented in section 7. This is an interesting technical problem to work out. 

Our results on the linearised higher spin super-Cotton tensors provide necessary prerequisites for developing a superspace approach 
to higher spin $\cN=2$ conformal supergravity.
We recall that the most general formulation\footnote{As explained in \cite{BKNT-M}, 
the conventional formulation for 3D $\cN$-extended conformal supergravity 
\cite{HIPT,KLT-M11}
is obtained from that given in \cite{BKNT-M} by partially fixing the gauge freedom.
In this sense, 3D $\cN$-extended conformal superspace of \cite{BKNT-M} 
is the   the most general formulation for 3D $\cN$-extended  conformal supergravity.}
for 3D $\cN$-extended conformal supergravity 
is the conformal superspace of \cite{BKNT-M}, which is a 3D analogue 
of the  4D conformal superspace formulations initiated by Butter \cite{ButterN=1,ButterN=2}.  
In this approach, it is the $\cN$-extended
super-Cotton tensor which fully determines the superspace geometry of conformal supergravity. 
It is necessary to mention that the program of constructing a superconformal theory of massless 
higher spin fields in (2+1) spacetime dimensions was put forward long ago by Fradkin and Linetsky
\cite{FL} in the component setting. However, it appears that superspace techniques
may offer new insights.

Our approach to constructing higher spin massive supermultiplets is a generalisation 
of topologically massive (super)gravity. Recently, there has appeared a conceptually different way to generate 3D massive (super)gravity theories -- new massive (super)gravity
theories \cite{BHT1,BHT2,Andringa:2009yc,BHRST10} 
and their generalisations, see  \cite{KNT-M,ABBOS} and references therein.
We believe that our results may be used to construct higher spin analogues of 
these massive theories.

Our massive transverse supermultiplet \eqref{eq:HSMassiveAction}
can be coupled to an external source $\cJ_{\a(2s)}$ using an action functional 
of the form 
\bea
 \m^{2s-1} S^{\perp}_{s+\frac{1}{2}} [H, \G, \bar \G] 
+\frac{\lambda}{2} S_{\rm CS} [H]
+\left(-\frac{1}{2}\right)^{s}\int \rd^{3|4}z \, H^{\a(2s)}\cJ_{\a(2s)} ~.
\label{8.1}
\eea
In order for such an action to be invariant under the  gauge transformations \eqref{eq:GTransHFlat} and \eqref{eq:GTransGammaFlat},
the real source $\cJ_{\a(2s)}$ must be conserved, that is
\bea
D^\b \cJ_{\b \a_1 \dots \a_{2s-1} } = \bar D^\b \cJ_{\b \a_1 \dots \a_{2s-1} } = 0~.
\eea
Such higher spin conserved current multiplets were considered in \cite{NSU}.
In  3D $\cN=2$ superconformal field theory,  $\cJ_{\a\b}$ describes 
the supercurrent multiplet \cite{DS,KT-M11}.\footnote{The two- and 
three-point functions of the $\cN=2$ supercurrent  were computed in \cite{BKS}.}
The theory with action \eqref{8.1} possesses a dual longitudinal formulation.
It is described by the action
\bea
 \m^{2s-1} S^{\parallel}_{s+\frac{1}{2}} [H, G, \bar G] 
+\frac{\lambda}{2} S_{\rm CS} [H]
 +\left(-\frac{1}{2}\right)^{s}\int \rd^{3|4}z \, H^{\a(2s)}\cJ_{\a(2s)} ~,
\label{8.3}
\eea
where the longitudinal action  $S^{\parallel}_{s+\frac{1}{2}} [H, G, \bar G] $
is given by eq. \eqref{eq:FlatAction3DLongitudinal}.
\\


\noindent
{\bf Acknowledgements:}\\
We are grateful to Mirian Tsulaia for comments on the manuscript. 
We acknowledge Gabriele Tartaglino-Mazzucchelli for discussions and collaboration 
 at the early stage of this work.
The work of SMK is supported in part by the Australian Research Council,
project No. DP160103633. The work of DXO is supported by
the Hackett Postgraduate Scholarship of The University of Western Australia.

\appendix

\section{Notation and conventions}

Our 3D notation and conventions correspond to those introduced in 
\cite{KPT-MvU,KLT-M11}. 
  
The spinor covariant derivatives have the form 
\begin{subequations}
\bea
D_{\alpha}&=& \frac{\pa}{\pa \q^\a}
+\ri \bar{\theta}^{\beta}(\g^a)_{\a\b} \pa_a
= \pa_\a + \ri \bar \q^\b \partial_{\alpha\beta}~, \\
\bar D_{\alpha}&=&-\frac{\pa}{\pa \bar \q^\a} -\ri {\theta}^{\beta}(\g^a)_{\a\b} \pa_a
=-\bar \pa_\a
-\ri {\theta}^{\beta}\partial_{\alpha\beta} 
\eea
\end{subequations}
and obey the anti-commutation relations 
\bea
\{ D_\a, D_\b \} = \{ \bar D_\a, \bar D_\b \} = 0~, \qquad
\{ D_\a, \bar D_\b \} = -2\ri \pa_{\a\b}~.
\eea
The generators of supersymmetry transformations are
\bea
Q_\a = \ri \pa_\a + \bar \q^\b \partial_{\alpha\beta}~, \qquad 
\bar Q_\a =-\ri \bar \pa_\a
- {\theta}^{\beta}\partial_{\alpha\beta} ~.
\eea
We make use of the definitions
\bea
D^2 = D^\a D_\a~, \qquad \bar D^2 = \bar D_\a \bar D^\a
\eea
such that the complex conjugate of $D^2 V$ is  $\bar D^2 \bar V$, for any superfield $V$.

Most tensor (super)fields encountered in this paper are completely symmetric 
with respect to their spinor indices. We use the rules introduced in \cite{Vasiliev}
and adopted in \cite{KS94}.
\begin{subequations}
\bea
V_{\a(n)} &=& V_{\a_1 \dots \a_n} = V_{(\a_1 \dots \a_n)} \ ,
\\
V_{ \a(n)} U_{\a(m) }&=& V_{ (\a_1 \dots \a_n} U_{\a_{n+1} \dots \a_{n+m})}\ , \\
V \cdot U &=& V^{\a(n) } U_{\a(n)} = V^{\a_1 \dots \a_n} U_{\a_1 \dots \a_n} \label{A.6} \ , \\
V_{(\a(n)} U_{\b(m) )} &=& V_{(\a_1 \dots \a_n} U_{\b_1 \dots \b_m )} \ .
\eea
\end{subequations}
Parentheses denote symmetrisation of indices.
Indices sandwiched between vertical  bars (e.g. $|\g|$) are not subject to symmetrisation. 
Throughout the entire paper, 
we assume that (super)fields carrying an even number of spinor indices correspond to bosons, 
whereas (super)fields carrying an odd number of spinor indices correspond to fermions.


\section{Massless higher spin actions in three dimensions}\label{AppendixB}

In this appendix we briefly review the 
(Fang-)Fronsdal actions 
for massless higher spin fields in three dimensions \cite{Fronsdal,FF}.
We also show that these models describe no propagating degrees
of freedom, as a simple extension of the 4D analysis in section 6.9 of \cite{BK}.

\subsection{Integer spin}\label{SectionB.1}

Given an integer $s>1$, we consider the following set of real bosonic  fields
\begin{equation}
\vf^i =
\Big\{h_{\a(2s)}\, ,  \,h_{\a(2s-4)}\Big\}
\label{B.1}
\end{equation}
defined modulo gauge transformations 
\begin{subequations} \label{B.2}
\bea
\delta h_{\a(2s)}&=&\pde_{(\a_{1}\a_{2}}\zeta_{\a_{3} \dots \a_{2s}) } ~,\ \label{B.2a}\\
\delta h_{\a(2s-4)}&=&
\frac{1}{2s-1}
\pde^{\b \g }\zeta_{\b \g \a_1 \dots \a_{2s-4} }~,
\label{B.2b}
\eea
\end{subequations}
where the gauge parameter 
$\zeta_{\a(2s-2)}$ is real. 
It may be checked that the following action 
\bea
S_s &=&\frac{1}{2}\Big(-\frac{1}{2}\Big)^{s}\int \rd^{3}x\,
\bigg\{ h^{\a(2s)}\Box h_{\a(2s)}
-\hf s \Big(\pde^{\b(2)}h_{\b(2)\a(2s-2)}\Big)^{2}
\nonumber \\
&&-(s-1)(2s-3)\bigg[ 
h^{\a(2s-4)}\pde^{\b(2)}\pde^{\gamma(2)}h_{\b(2)\gamma(2)\a(2s-4)}
+4\frac{(s-1)}{s} h^{\a(2s-4)}\Box h_{\a(2s-4)}
\non \\
&&
+\hf (s-2)(2s-5)\Big(\pde^{\b(2)}h_{\a(2s-6)\b(2)}\Big)^{2} \bigg]\bigg\}
\label{B.3}
\eea
is gauge invariant. The requirement of gauge invariance determines the action 
uniquely modulo an overal constant. 
The theory admits a formal limit to the case of three-dimensional Maxwell's electrodynamics
It is obtained by setting $s=1$, removing the second field in \eqref{B.1}, and switching 
off all the terms in the second and third lines of the action \eqref{B.3}.

The equations of motion are:
\begin{subequations} \label{B.4}
\bea
&&\Box h_{\a(2s)}
+\hf s\pde^{\b(2)}\pde_{\a(2)}h_{\a(2s-2)\b(2)}
-\hf (s-1)(2s-3)\pde_{\a(2)}\pde_{\a(2)} h_{\a(2s-4)}=0~, ~~~~\label{B.4a}\\
&&\pde^{\b(2)}\pde^{\gamma(2)}h_{\b(2)\gamma(2)\a(2s-4)}
+8 \frac{(s-1)}{s}\Box h_{\a(2s-4)} \non \\
&&\qquad \qquad \qquad 
+(s-2)(2s-5) \pde^{\b(2)}\pde_{\a(2)}h_{\a(2s-6)\b(2)}=0~.~~~~ \label{B.4b}
\eea
\end{subequations}
We now show that the model under consideration has no propagating degrees of freedom.

The gauge freedom \eqref{B.2} allows us to gauge away the field $h_{\a(2s-4)}$,
\bea
h_{\a(2s-4)}=0~.
\label{B.5}
\eea
In this gauge, there still remains a residual gauge freedom. In accordance with 
\eqref{B.2b}, the gauge parameter is now constrained by 
\bea
\pde^{\b (2) }\zeta_{\b (2) \a(2s-4) } =0~.
\label{B.6}
\eea
In the gauge \eqref{B.5}, the equation of motion \eqref{B.4b} reduces to 
$\pde^{\b(2)}\pde^{\gamma(2)}h_{\b(2)\gamma(2)\a(2s-4)} =0$ and 
tells us that $\pde^{\b(2)} h_{\a(2s-2) \b(2)}$ is divergenceless. 
In general, it holds that 
\bea
\pde^{\b(2)} \d h_{\a(2s-2) \b(2)} =-\frac{2}{s} \Box\z_{\a(2s-2)} 
+\frac{(s-1)(2s-3)}{
s(2s-1)} \pa^{\b(2)} \pa_{\a(2)} \z_{\a(2s-4) \b(2)}~.
\eea
Under the two conditions that (i)  the gauge parameter is constrained as in \eqref{B.6},
and (ii) $\pde^{\b(2)} h_{\a(2s-2) \b(2)}$ is divergenceless, we are able to impose the 
gauge condition 
\bea
\pde^{\b(2)} h_{\a(2s-2) \b(2)} =0~,
\label{B.8}
\eea
in addition to \eqref{B.5}. The residual gauge freedom, which respects the conditions
\eqref{B.5} and \eqref{B.8}, is generated by a gauge parameter constrained by 
\bea
\pde^{\b (2) }\zeta_{\b (2) \a(2s-4) } =0~, \qquad \Box \z_{\a(2s-2)}=0~.
\label{B.9}
\eea
Due to \eqref{B.5} and \eqref{B.8}, the equation of motion \eqref{B.4a} 
turns into 
\bea
\Box h_{\a(2s)}=0~.
\eea

Since both the field $h_{\a(2s)} (x) $ and the gauge parameter $\z_{\a(2s-2)}(x)$
are on-shell, it is useful to switch to momentum space, by replacing 
$h_{\a(2s)} (x) \to h_{\a(2s)} (p)  $ and $\z_{\a(2s-2)}(x) \to\z_{\a(2s-2)}(p)$, where
the three-momentum $p^a$ is light-like, $p^{\a\b}p_{\a\b}=0$. For a given three-momentum,
we can choose a frame in which the only non-zero component of 
$p^{\a\b}= (p^{11}, p^{12} =p^{21}, p^{22}) $ is $p^{22}=p_{11}$. 
Then, the conditions $p^{\b(2)}h_{\a(2s-2) \b(2)} (p)=0$ and 
$p^{\b(2)}\z_{\a(2s-4) \b(2)} (p)=0$ are equivalent to 
\bea
h_{\a(2s-2) 22} (p)=0~, \qquad \z_{\a(2s-4) 22} (p)=0~.
\eea
We see that $h_{\a(2s)} $ has only two independent components, 
which are: $h_{1\dots 1}$ and $h_{1\dots12}$, and similar for 
the gauge parameter $\z_{\a(2s-4)}$. The gauge transformation law 
\eqref{B.2a} now amounts to $\d h_{1\dots 1} \propto p_{11} \z_{1\dots1}$
and $\d h_{1\dots 12} \propto p_{11} \z_{1\dots12}$. As a result, 
the field $h_{\a(2s)} $ can be completely gauged away for $s>1$.
The case $s=1$ is special. Here the field $h_{\a\b}$ has again two components,
$h_{11}$ and $h_{12}$, while the gauge parameter is a scalar, $\z$. 
The latter allows us to gauge away $h_{11}$, since its gauge transformation is
$\d h_{11} \propto p_{11} \z$. The other component, $h_{12}$, describes a propagating degree 
of freedom. In the gauge $h_{11}=0$, it is proportional to a single non-zero component of 
the gauge-invariant field strength $F^a = \hf \ve^{aba}F_{bc} $, where 
$F_{ab} = \pa_a h_b -\pa_b h_a$.

\subsection{Half-integer spin} \label{AppendixB.2}

Given an integer $s>1$, we consider the following set of real fermionic  fields
\begin{equation}
\phi^{{j}}=\Big\{\j_{\a(2s+1)},\j_{\a(2s-1)},\psi_{\a(2s-3)} \Big\}
\end{equation}
defined modulo gauge transformations of the form
\begin{subequations} \label{B.13}
\bea
\delta \j_{\a(2s+1)}&=&\pde_{(\a_1\a_2 }\xi_{\a_3 \dots \a_{2s+1} )} ~,
\label{B.13a}\\
\delta \j_{\a(2s-1)}&=&\frac{2s-1}{2s+1}\pde^\b{}_{(\a_1}\xi_{\a_2 \dots \a_{2s-1})\b}~,
\label{B.13b}\\
\delta \psi_{\a(2s-3)}&=&
\pde^{\b \g}\xi_{\a_1 \dots \a_{2s-3} \b \g}~, \label{B.13c}
\eea
\end{subequations}
where the gauge parameter $\x_{\a(2s-1)}$ is real. 
It may be checked that the following action 
\bea
S_{s+\hf}&=& \frac{\ri}{2} \Big(\!-\frac{1}{2}\Big)^{s}\int \rd^{3}x \,
\bigg\{\j^{\a(2s)\b}\pde_\b{}^{\gamma} \j_{\g \a(2s)}
+2 \j^{\a(2s-1)}\pde^{\b(2)}\j_{\b(2) \a(2s-1)}\nonumber \\
&&+\frac{4}{2s-1}\j^{\a(2s-2)\b}\pde_\b{}^{\gamma}{\j}_{\g\a(2s-2)} 
 \label{B.14} \\
&&+\frac{(s-1)(2s+1)}{s(2s-1)} \Big( 
2 {\psi}^{\a(2s-3)}\pde^{\b(2)}\phi_{\b(2) \a(2s-3)}
-\frac{2s-3}{2s+1} {\psi}^{\a(2s-4)\b}\pde_\b{}^{\gamma}\psi_{\gamma\a(2s-4)} \Big) 
\bigg\} ~~~~~
\non 
\eea
is gauge invariant. The field $\psi_{\a(2s-3)} $ is not defined in the case $s=1$ 
which corresponds to the massless gravitino. However, the last line in \eqref{B.14}, 
which contains all the dependence on $\psi_{\a(2s-3)} $,
does not contribute in the case, due to the overall factor of $(s-1)$.
Thus the gravitino action follows from \eqref{B.14} by deleting the third line and 
then setting $s=1$. 

The equations of motion are:
\begin{subequations} \label{B.15}
\bea
&&
\pde^\b{}_{\a}\j_{\a(2s)\b}-\pde_{\a(2)}\j_{\a(2s-1)} =0~,  \label{B.15a}
\\
&& \pde^{\b(2)}\j_{\a(2s-1)\b(2)}+
\frac{4}{2s-1}\Big[
\pde^\b{}_{\a} \j_{\a(2s-2)\b}
-
\frac{(s-1)(2s+1)}{s} \pde_{\a(2) }\psi_{\a(2s-3)}\Big]=0~,~~~~~~~
 \label{B.15b}
\\
&&\pde^{\b(2)}\j_{\a(2s-3)\b(2)}-\frac{2s-3}{2s+1}
\pde^\b{}_{\a} \psi_{\a(2s-4)\b} =0
~.  \label{B.15c}
\eea
\end{subequations}
We now show that the model under consideration has no propagating degrees of freedom.

The gauge freedom \eqref{B.13c} allows us to gauge away the field $\j_{\a(2s-3)}$,
\bea
\j_{\a(2s-3)} =0~.
\label{B.16}
\eea
In this gauge, there still remains a residual gauge freedom. In accordance with 
\eqref{B.13c}, the gauge parameter is now constrained by 
\bea
\pde^{\b (2) }\x_{\b (2) \a(2s-3) } =0~.
\label{B.17}
\eea
In the gauge \eqref{B.16}, the equation \eqref{B.15c} reduces to 
\bea
\pde^{\b(2)}\j_{\a(2s-3)\b(2)} =0~,
\label{B.18}
\eea
which is preserved by the residual gauge transformations, 
 as a consequence of \eqref{B.17}.
 Due to \eqref{B.17} and \eqref{B.18}, it follows from the gauge transformation 
 \eqref{B.13b} that the field $\j_{\a(2s-1)}$ may be gauged away, 
 \bea
 \j_{\a(2s-1)}=0~.
 \label{B.19}
 \eea
Under this gauge condition, there still remains some residual gauge freedom. 
It is described by an on-shell  parameter $\x_{\a(s2-1)}$, which is  constrained by 
\bea
\pde^\b{}_{\a_1} \x_{\a_2 \dots \a_{2s-1} \b}=0 \quad \Longrightarrow
\quad \Box \x_{\a(2s-1)}=0~,
\label{B.20}
\eea
in addition to \eqref{B.17}. Under the gauge conditions \eqref{B.16} and \eqref{B.19}, 
the equations of motion \eqref{B.15} amount to 
\bea
\pde^\b{}_{\a_1} \j_{\a_2 \dots \a_{2s+1} \b}=0 \quad \Longrightarrow
\quad \pa^{\b(2)} \j_{\a(2s-1) \b(2)} =0~, \qquad 
\Box \j_{\a(2s+1)}=0~.
\eea

Since both the field $j_{\a(2s+1)} (x) $ and the gauge parameter $\x_{\a(2s-1)}(x)$
are on-shell, it is useful to switch to momentum space, by replacing 
$\j_{\a(2s+1)} (x) \to \j_{\a(2s+1)} (p)  $ and $\x_{\a(2s-1)}(x) \to\x_{\a(2s-1)}(p)$, where
the three-momentum $p^a$ is light-like, $p^{\a\b}p_{\a\b}=0$. 
As in the bosonic case studied in the previous subsection,  
we can choose a frame in which the only non-zero component of 
$p^{\a\b}= (p^{11}, p^{12} =p^{21}, p^{22}) $ is $p^{22}=p_{11}$. 
Then, the conditions $p^\b{}_{\a_1} \j_{\a_2 \dots \a_{2s+1} \b}(p)=0$ and 
$p^\b{}_{\a_1} \x_{\a_2 \dots \a_{2s-1} \b}(p)=0$ are equivalent to 
\bea
\j_{\a(2s)2 }(p)=0~, \qquad \x_{\a(2s-2)2 }(p)=0~.
\eea
Thus the only non-zero components of $\j_{\a(2s+1)} (p)$ and  $\x_{\a(2s-1)} (p)$ 
are $\j_{1\dots 1}(p) $ and  $\x_{1\dots 1}(p) $. 
The residual gauge freedom, 
$\d \j_{\a (2s+1) } (p) \propto p_{\a (2) } \x_{\a(2s-1)}$ 
allows us to gauge away the field $\j_{\a(2s+1) } $ completely. 
 A minor modification of the above analysis can be used in the case $s=1$ to show 
 that the massless gravitino action does not describe any propagating degrees of freedom. 
 

\section{Component reduction}\label{AppendixC}  
  
Here we shall elaborate on the component structure of the massless 
superspin-$(s+\hf)$  model in the transverse formulation \eqref{eq:FlatAction3DTrans}. 
The longitudinal action  \eqref{eq:FlatAction3DLongitudinal} can be reduced to 
components in a similar fashion. Our approach to the component reduction 
of  \eqref{eq:FlatAction3DTrans} will be similar to that used in \cite{KSP,KS93,KS94}
for the off-shell higher spin $\cN=1$ supermultiplets 
in four dimensions.\footnote{Further aspects of the component structure 
of the off-shell higher spin models proposed in  \cite{KSP,KS93} were studied in 
\cite{GatesK}.}

 It is useful to define the components fields of a superfield using the standard bar-projection
 \bea
 U|:= U(x, \q, \bar \q) |_{\q = \bar \q=0}~,
 \eea
 for any superfield $U(z)$.
 Our definition of the component fields of $H_{\a(2s)}$ and $\Gamma_{\a(2s-2)}$ 
 will be  consistent with the Wess-Zumino gauge  \eqref{eq:WZGauge_new}. 

In the Wess-Zumino gauge  \eqref{eq:WZGauge_new},
the component fields of $H_{\a(2s)}$ are:
\begin{subequations}
 \begin{align}
 h_{\a(2s+2)}:=&\frac{1}{2}[D_{(\a_{1}},\bar{D}_{\a_{2}}]H_{\a_{3}..\a_{2s+2})}\vert
 =D_{(\a_{1}}\bar{D}_{\a_{2}}H_{\a_{3}...\a_{2s+2})}\vert ~, &\\
 \Psi_{\a(2s+1)}:=&-\frac{1}{4}\bar{D}^{2}D_{(\a_1}H_{\a_2 \dots \a_{2s+1})}\vert=-\frac{1}{4}D_{(\a_1}\bar{D}^{2}H_{\a_2 \dots \a_{2s+1})}\vert ~, &\\
 A_{\a(2s)}:=& \frac{1}{32}\{D^{2},\bar{D}^{2}\}H_{\a(2s)}\vert ~.&
 \end{align}
 \end{subequations}
  The component fields of
 $\Gamma_{\a(2s-2)}$ are: 
\begin{subequations}
 \begin{align}
 \gamma_{\a(2s-2)}:=&\Gamma_{\a(2s-2)}\vert=\bar{\gamma}_{\a(2s-2)} ~, &\\
 \Psi_{\a(2s-1)}:=&+D_{( \a_1}\Gamma_{\a_2 \dots \a_{2s-1})}\vert ~, &\\
 \bar{\Psi}_{\a(2s-3)}:=&-\frac{2s-2}{2s-1}D^{\b}\Gamma_{\a(2s-3)\b}\vert ~, &\\
 \U_{\a(2s-1)}:=&-\bar{D}_{(\a_1}\Gamma_{\a_2 \dots \a_{2s-1})}\vert ~, &\\
 B_{\a(2s-2)}:=&-\frac{1}{4}D^{2}\Gamma_{\a(2s-2)}\vert ~, &\\
 U_{\a(2s)}:=&+\frac{1}{2}[D_{(\a_{1}},\bar{D}_{\a_{2}}]\Gamma_{\a_{3}...\a_{2s})}\vert ~, &\\
 F_{\a(2s-2)}:=& \frac{2s-1}{2s}D^{\b}\bar{D}_{\b}\Gamma_{\a(2s-2)}\vert ~, &\\
 \rho_{\a(2s-1)}:=&\frac{1}{8}D^{2}\bar{D}_{(\a_1}\Gamma_{\a_2 \dots \a_{2s-1})} \vert ~.&
 \end{align}
 \end{subequations}
Introducing the superfield Lagrangian
 $ \mathcal{L}^{\perp}_{s+\frac{1}{2}}  $ for the transverse action  \eqref{eq:FlatAction3DTrans}, 
\bea
S^{\perp}_{s+\frac{1}{2}}[H,\G, \bar \G ]= \int \rd^{3|4}z \,\mathcal{L}^{\perp}_{s+\frac{1}{2}} 
\eea
the component Lagrangian $L$ is defined by 
\begin{align}
S^{\perp}_{s+\frac{1}{2}}[H,\G, \bar \G ]
=\frac{1}{16}  \int \rd^{3}x \,D^{2}\bar{D}^{2}
\mathcal{L}^{\perp}_{s+\frac{1}{2}}\Big|
= \int \rd^{3}x \,{L}~.
\end{align}
The component Lagrangian
naturally splits into its bosonic and fermionic parts:
\begin{align}
 L = L_{\rm bos} +L_{\rm ferm}~.
\end{align}
Below we analyse separately the bosonic and fermionic sectors of $L$.


\subsection{Bosonic sector}

For the bosonic Lagrangian we obtain 
\begin{align}
L_{\rm bos}=&(-\frac{1}{2})^{s} \bigg\{2A \cdot A+\frac{1}{8}(\pde \cdot h)^{2}
-\frac{1}{4}h^{\a(2s+2)}\Box h_{\a(2s+2)}\nonumber &\\
&+A\cdot (U+\bar{U})+\frac{1}{4}\mri (\pde \cdot h)\cdot (U-\bar{U})-\frac{1}{2} (\pde \cdot \pde \cdot h)^{\a(2s-2)}\gamma_{\a(2s-2)}\nonumber &\\
&-\frac{2s^{2}-5s+1}{s^{2}}\gamma^{\a(2s-2)}\cdot \Box \gamma_{\a(2s-2)} -\frac{(s-1)(2s-3)(4s-1)}{2s^{2}(2s-1)}(\pde^{\b(2)} \gamma_{\a(2s-4)\b(2)})^{2} \nonumber &\\
&-\frac{1}{2(2s-1)}F\cdot \bar{F}-\mri\frac{s-1}{s}\gamma^{\a(2s-3)\rho}\pde_{\rho}{}^{\b}(F-\bar{F})_{\a(2s-3)\b} & \non \\
&
+\frac{(2s+1)}{4(2s-1)}(F\cdot F +\bar{F} \cdot \bar F)\bigg\}~,&
\end{align}
where the dot notation $F\cdot \bar{F}$ and $\pde \cdot h$ stands for the  contraction of spinor indices, 
for instance:  $F^{\a(2s-2)}\bar{F}_{\a(2s-2)}$ and $\pde^{\b(2)}h_{\a(2s)\b(2)}$. 
Integrating out the auxiliary fields $F_{\a(2s-2)}, U_{\a(2s-2)}, A_{\a(2s)}$ and $B_{\a(2s)}$, we arrive at the following Lagrangian:
\begin{align}
\hat{L}_{\rm bos}
=&(-\frac{1}{2})^{s}\bigg\{-\frac{1}{4}h^{\a(2s+2)}\Box h_{\a(2s+2)} 
+\frac{s+1}{8}(\pde \cdot h)^{2}
+\frac{2s-1}{2} (\pde \cdot \pde \cdot h)^{\a(2s-2)}\gamma_{\a(2s-2)}\nonumber &\\
&+\frac{4(2s-1)}{s+1}\gamma\cdot \Box \gamma+\frac{(s-1)(2s-1)(2s-3)}{2s(s+1)}(\pde \cdot \gamma)^{2}\bigg\}~.&
\label{C.8}
\end{align}

The gauge transformations of the component fields $h_{\a(2s+2)}$ and $h_{\a(2s-2)}$ can be read from the gauge transformations of the $H_{\a(2s)}$ and $\Gamma_{\a(2s-2)}$ superfields, respectively, 
 in terms of the longitudinal linear gauge parameter $g_{\a(2s)}$. 
  In the Wess-Zumino gauge, we have
\begin{equation}
\delta h_{\a(2s+2)}= \pde_{(\a_1\a_2}\z_{\a_3 \dots \a_{2s+2})}~, \qquad 
\delta \gamma_{\a(2s-2)}= \frac{s}{2(2s+1)}\pde^{\b\g}  
\z_{\a(2s-2)\b \g} ~.\label{eq: C.9}
\end{equation}
We recall that the real gauge parameter $\z_{\a(2s)}(x) $ originates as 
$g_{\a(2s)}\vert=-\frac{\ri}{2} \z_{\a(2s)}$, see eq. \eqref{com5.4a}.

We now compare \eqref{C.8} with the Lagrangian 
corresponding to the massless action given in section 
\ref{SectionB.1} with spin $s$ replaced with  $s+1$:
\bea
S_{s+1} &=&\frac{1}{2}\Big(-\frac{1}{2}\Big)^{s+1}\int \rd^{3}x\,
\bigg\{ h^{\a(2s+2)}\Box h_{\a(2s+2)}
-\frac{s+1}{2} \Big(\pde^{\b(2)}h_{\b(2)\a(2s)}\Big)^{2}
\nonumber \\
&&-s(2s-1)\bigg[ 
h^{\a(2s-2)}\pde^{\b(2)}\pde^{\gamma(2)}h_{\b(2)\gamma(2)\a(2s-4)}
+4\frac{s}{s+1} h^{\a(2s-2)}\Box h_{\a(2s-2)}
\non \\
&&
+\hf (s-1)(2s-3)\Big(\pde^{\b(2)}h_{\a(2s-4)\b(2)}\Big)^{2} \bigg]\bigg\}.
\label{C.10}
\eea
Clearly, the Lagrangians coincide if we make the identification 
\begin{equation}
\gamma_{\a(2s-2)}=(2s+1)h_{\a(2s-2)}.
\end{equation}
In this manner, all terms in the bosonic sector agree with Fronsdal's action.


\subsection{Fermionic sector}

For the fermionic Lagrangian we obtain
\begin{align}
L_{\rm ferm}
=&\mri(-\frac{1}{2})^{s} \bigg\{\Psi^{\a(2s)\b}\pde_{\b}{}^{\gamma}\bar{\Psi}_{\a(2s)\gamma}\nonumber &\\
&+\Big(\bar{\Psi}^{\a(2s+1)}\pde_{(\a_{2s+1}\a_{2s}}\U_{\a(2s-1))}+\Psi^{\a(2s+1)} \cdot \pde_{\a_{2s+1}\a_{2s}}\bar{\U}_{\a(2s-1))}\Big)\nonumber &\\
&-\frac{2s-1}{2s\mri}\Big(\rho^{\a(2s-1)}\bar{\Psi}_{\a(2s-1)}-\bar{\rho}^{\a(2s-1)}\Psi_{\a(2s-1)} \Big)\nonumber &\\
&+\frac{2s-1}{2s} \Psi^{\a(2s-2)\b}\pde_{\b}{}^{\rho}\bar{\Psi}_{\a(2s-2)\rho}+\frac{2s-1}{2s} \Psi^{\a(2s-1)}\pde_{\a_{2s-1}\a_{2s-2}}\Psi_{\a(2s-3)}\nonumber&\\
&-\frac{2s-1}{2s} \bar{\Psi}^{\a(2s-3)}(\pde \cdot \bar{\Psi})_{\a(2s-3)}- \frac{2s-3}{2s-2}\frac{2s-1}{2s}\bar{\Psi}^{\a(2s-4)\b}\pde^{\rho}{}_{\b}\Psi_{\a(2s-4)\rho}\nonumber &\\
&-\frac{2s-1}{2s} \U^{\a(2s-2)\b}\pde_{\b}{}^{\gamma}\bar{\U}_{\a(2s-2)\gamma}\nonumber &\\
&+\frac{2s+1}{2s\mri}\Big(\rho^{\a(2s-2)\b}\U_{\a(2s-2)\b}-\bar{\rho}^{\a(2s-2)\b}\bar{\U}_{\a(2s-2)\b}\Big)\bigg\}~.&
\end{align}
The fields $\rho_{\a(2s-1)}$ and $\U_{\a(2s-1)}$ are auxiliary.
Integrating them out leads to the following Lagrangian 
involving only the dynamical fields $\Psi_{\a(2s+1)}$, $\Psi_{\a(2s-1)}$
and  $\Psi_{\a(2s-3)}$: 
\begin{align}
\hat{L}_{\rm ferm}
=&\mri(-\frac{1}{2})^{s} \bigg\{\Psi^{\a(2s)\b}\pde_{\b}{}^{\gamma}\bar{\Psi}_{\a(2s)\gamma}-\frac{2s-1}{2s+1}\Big(\bar{\Psi}^{\a(2s-1)}\pde^{\b(2)}\bar{\Psi}_{\a(2s-1)\b(2)}+{\rm c.c.}\Big)\nonumber &\\ 
&+\frac{4(2s-1)}{(2s+1)^{2}}\Psi^{\a(2s-2)\b}\pde_{\b}{}^{\rho}\bar{\Psi}_{\a(2s-2)\rho}\nonumber &\\
&-\frac{2s-1}{2s}(\bar{\Psi}^{\a(2s-3)}(\pde \cdot \bar{\Psi})_{\a(2s-3)}+\Psi^{\a(2s-3)}(\pde \cdot \Psi)_{\a(2s-3)})&\non \\
&- \frac{2s-3}{2s-2}\frac{2s-1}{2s}\bar{\Psi}^{\a(2s-4)\b}\pde^{\rho}{}_{\b}\Psi_{\a(2s-4)\rho}\bigg\}~.&
\end{align}
It is useful to perform the rescaling 
$$
\Big\{\Psi_{\a(2s+1)}, \Psi_{\a(2s-1)}, \Psi_{\a (2s-3)} \Big\} \rightarrow 
\Big\{\Psi_{\a(2s+1)},-\Psi_{\a(2s-1)},\frac{2(s-1)}{2s-1}\Psi_{\a (2s-3)}\Big\}~,$$ 
which leads to a massless fermionic  action
\begin{align}
\widetilde{S}_{\rm ferm}=& \mri(-\frac{1}{2})^{s}\int \rd^{3}x \,\bigg\{\bar{\Psi}^{\a(2s)\b}\pde^{\gamma}{}_{\b}\Psi_{\a(2s)\gamma}
+\frac{2s-1}{2s+1}(\bar{\Psi}^{\a(2s-1)}\pde^{\b(2)}\Psi_{\a(2s-1)\b(2)}+{\rm c.c.})\nonumber &\\
&+\frac{4(2s-1)}{(2s+1)^{2}}\Psi^{\a(2s-2)\b}\pde^{\gamma}{}_{\b}\bar{\Psi}_{\a(2s-2)\gamma}+\frac{s-1}{s}(\bar{\Psi}^{\a(2s-3)}\pde^{\b(2)}\Psi_{\a(2s-3)\b(2)}+{\rm c.c.})\nonumber &\\
&-\frac{2s-3}{2s-1}\frac{s-1}{s}\bar{\Psi}^{\a(2s-4)\b}\pde^{\gamma}{}_{\b}\Psi_{\gamma\a(2s-4)}\bigg\} \label{eq:FermionicFronsdalComplex}~,&
\end{align}
which proves to be invariant under  gauge transformations of the form
\begin{subequations}
\begin{align}
\delta \Psi_{\a(2s+1)}=&\pde_{(\a_1\a_2}\xi_{\a_3 \dots \a_{2s+1})}~,&\\
\delta \Psi_{\a(2s-1)}=&\pde^\b{}_{(\a_1} \xi_{\a_2 \dots \a_{2s-1})\b} ~,&\\
\delta \Psi_{\a(2s-3)}=&\pde^{\b\g}\xi_{\a(2s-3)\b\g}~,&
\end{align}
\end{subequations}
with $\xi_{\a(2s-1)}$ a complex gauge parameter. 

Note that the action \eqref{eq:FermionicFronsdalComplex} involving the complex fields $\{\Psi_{\a(2s+1)},\Psi_{\a(2s-1)},\Psi_{\a(2s-3)}\}$ is a sum of two actions 
involving real fields. More precisely, let us define the action 
\begin{align}
S_{s+\frac{1}{2}}^{J}:=& \frac{\mri}{2}(-\frac{1}{2})^{s}\int \rd^{3}x \,\bigg\{\psi^{J\,\a(2s)\b}\pde^{\gamma}{}_{\b}\psi_{\a(2s)\gamma}^{J}
+2\frac{2s-1}{2s+1}\psi^{J\, \a(2s-1)}\pde^{\b(2)}\psi_{\a(2s-1)\b(2)}^{J}\nonumber &\\
&+\frac{4(2s-1)}{(2s+1)^{2}}\psi^{J\,\a(2s-2)\b}\pde^{\gamma}{}_{\b}\psi_{\a(2s-2)\gamma}^{J}
+\frac{s-1}{s}\psi^{J\,\a(2s-3)}\pde^{\b(2)}\psi_{\a(2s-3)\b(2)}^{J} \nonumber &\\
&-\frac{2s-3}{2s-1}\frac{s-1}{s}{\psi}^{J\,\a(2s-4)\b}\pde^{\gamma}{}_{\b}\psi_{\gamma\a(2s-4)}^{J}\bigg\}~,
 \label{eq:FermionicFronsdalReal}&
\end{align}
which is invariant under the following gauge symmetry 
\begin{subequations}
\begin{align}
\delta \psi_{\a(2s+1)}^{J}=&\pde_{(\a_{1}\a_{2}}\xi_{\a_3 \dots \a_{2s+1})} ~,&\\
\delta \psi_{\a(2s-1)}^{J}=&\pde^\b{}_{(\a_1} \xi_{\a_2 \dots \a_{2s-1})\b} ~,&\\
\delta \psi_{\a(2s-3)}^{J}=&\pde^{\b \g}\xi_{\a(2s-3)\b \g}~,&
\end{align}
\end{subequations}
with real gauge parameter $\xi_{\a(2s-1)}$. 
Here
the label $J=\1,\2$ denotes two identical sets of real fermionic fields, $\{\psi^{J}_{\a(2s+1)},\psi^{J}_{\a(2s-1)},\psi^{J}_{\a(2s-3)}\}$. 
Each action \eqref{eq:FermionicFronsdalReal} is equivalent to the massless 
spin-$(s+\hf)$ model \eqref{B.14}, and the precise identification is as follows: 
\bea
\psi^{J}_{\a(2s+1)} = \j_{\a(2s+1)}~, \quad
\psi^{J}_{\a(2s-1)} =\frac{2s+1}{2s-1}\j_{\a(2s-1)}~,\quad \psi^{J}_{\a(2s-3)} =\j_{\a(2s-3)}~.
\eea 
It follows then that
\begin{equation}
\widetilde{S}_{\rm ferm}
=S_{s+\frac{1}{2}}^{\1}+S_{s+\frac{1}{2}}^{\2}~,
\end{equation}
with the complex  fields $\J$ related to the real ones $\j^J$
by the rule:
\begin{subequations}
\begin{align}
\sqrt{2} \Psi_{\a(2s+1)}=&\psi^{\1}_{\a(2s+1)}+\mri \psi^{\2}_{\a(2s+1)} ~,&\\
\sqrt{2}\Psi_{\a(2s-1)}=&\psi^{\1}_{\a(2s-1)}-\mri \psi^{\2}_{\a(2s-1)} ~,&\\
\sqrt{2}\Psi_{\a(2s-3)}=& \psi^{\1}_{\a(2s-3)}-\mri \psi^{\2}_{\a(2s-3)}~.&
\end{align}
\end{subequations}


\section{Properties of the superconformal field strength}\label{AppendixD}

In this appendix we prove (i)  invariance of 
the field strength \eqref{eq:HSFSUniversal} under the gauge transformation 
\eqref{5.19}; and (ii) the Bianchi identies \eqref{6.244}.

\subsection{Gauge invariance}

The superconformal field strength \eqref{eq:HSFSUniversal} 
is constructed from the superfields
\begin{subequations}\label{eq:HSFSBasis}
\begin{align}
X^{J}_{\a(n)}:=&\Delta \Box^{J}\pde_{(\a_{1}}{}^{\b_{1}}\pde_{\a_{2}}{}^{\b_{2}}...\pde_{\a_{n-2J}}{}^{\b_{n-2J}}H_{\a_{n-2J+1}...\a_{n})\b(n-2J)}~, &\\
Z^{J}_{\a(n)}:=&\Delta^{2}\Box^{J}\pde_{(\a_{1}}{}^{\b_{1}}\pde_{\a_{2}}{}^{\b_{2}}...\pde_{\a_{n-1-2J}}{}^{\b_{n-1-2J}}H_{\a_{n-2J}...\a_{n})\b(n-1-2J)}~.
 &
\end{align}
\end{subequations}
which form a basis for the field strength. 
In order to prove  the gauge invariance of the field strength, 
it suffices  to vary
\eqref{eq:HSFSBasis} and then impose the condition of gauge invariance. 
The condition of gauge invariance generates a recursion relation which is solved 
to give the binomial coefficients appearing in \eqref{eq:HSFSUniversal}. Alternatively, one can directly show that $W_{\a(n)}$ is gauge invariant as follows. 
First we compute the variations of the basis superfields \eqref{eq:HSFSBasis}.
The results are:
\begin{subequations}
\begin{align}
\delta_{L} X^{J}_{\a(n)}:=&\Delta \Box^{J}\pde_{(\a_{1}}{}^{\b_{1}}\pde_{\a_{2}}{}^{\b_{2}}...\pde_{\a_{n-2J}}{}^{\b_{n-2J}}\delta H_{\a_{n-2J+1}...\a_{n})\b(n-2J)}
\nonumber &\\
=&\Delta\bigg\{\frac{J}{n/2}\Box^{J}\pde_{(\a_{1}}{}^{\b_{1}}\pde_{\a_{2}}{}^{\b_{2}}...\pde_{\a_{n-2J}}{}^{\b_{n-2J}}\bar{D}_{\a_{n-2J+1}}L_{...\a_{n})\b(n-2J)}\nonumber &\\
&+\frac{n/2-J}{n/2}\Box^{J}\pde_{(\a_{1}}{}^{\b_{1}}\pde_{\a_{2}}{}^{\b_{2}}...\pde_{\a_{n-2J}}{}^{\b_{n-2J}}\bar{D}_{|\b_{1}}L_{\b_{2}...\b_{n-2J}|\a_{n-2J+1}...\a_{n})}\bigg\}
\nonumber &\\
=& \mri\bigg\{\frac{J}{n} \Box^{J}\pde_{(\a_{1}}{}^{\b_{1}}\pde_{\a_{2}}{}^{\b_{2}}...\pde_{\a_{n-2J}}{}^{\b_{n-2J}}D_{\a_{n-2J+1}}\bar{D}^{2}L_{...\a_{n})\b(n-2J)}\nonumber &\\
&+\frac{n-2J}{2n}\Box^{J}\pde_{(\a_{1}}{}^{\b_{1}}\pde_{\a_{2}}{}^{\b_{2}}...\pde_{\a_{n-2J}}{}^{\b_{n-2J}}D_{|\b_{1}}\bar{D}^{2}L_{\b_{2}...\b_{n-2J}|\a_{n-2J+1}...\a_{n})}\bigg\}~, \nonumber &\\ 
 &\hspace{5cm} 0\leq J \leq \floor{n/2} ~;&\\
\delta_{L} Z^{J}_{\a(n)}:=&\Delta^{2}\Box^{J}\pde_{(\a_{1}}{}^{\b_{1}}\pde_{\a_{2}}{}^{\b_{2}}...\pde_{\a_{n-1-2J}}{}^{\b_{n-1-2J}}\delta H_{\a_{n-2J}...\a_{n})\b(n-1-2J)}
\nonumber &\\
=& \Delta \bigg\{-\frac{n-(2J+1)}{n/2}\Box^{J+1}\pde_{(\a_{1}}{}^{\b_{1}}\pde_{\a_{2}}{}^{\b_{2}}...\pde_{\a_{n-2J-2}}{}^{\b_{n-2J-2}}\bar{D}_{\a_{n-2J-1}}L_{\a_{n-2J}...\a_{n})\b(n-2J-2)}\nonumber&\\
&-\frac{2J+1}{n/2}\Box^{J}\pde_{(\a_{1}}{}^{\b_{1}}\pde_{\a_{2}}{}^{\b_{2}}...\pde_{\a_{n-2J}}{}^{\b_{n-2J}}\bar{D}_{|\b_{1}}L_{\b_{2}...\b_{n-2J}|\a_{n-2J+1}...\a_{n})}\bigg\}
\nonumber &\\
=&  \mri\bigg\{-\frac{n-(2J+1)}{n} \Box^{J+1}\pde_{(\a_{1}}{}^{\b_{1}}\pde_{\a_{2}}{}^{\b_{2}}...\pde_{\a_{n-2J-2}}{}^{\b_{n-2J-2}}D_{\a_{n-2J-1}}\bar{D}^{2}L_{\a_{n-2J}...\a_{n})\b(n-2J-2)}\nonumber&\\
&-\frac{2J+1}{n}\Box^{J}\pde_{(\a_{1}}{}^{\b_{1}}\pde_{\a_{2}}{}^{\b_{2}}...\pde_{\a_{n-2J}}{}^{\b_{n-2J}}D_{|\b_{1}}\bar{D}^{2}L_{\b_{2}...\b_{n-2J}|\a_{n-2J+1}...\a_{n})}\bigg\}~, \non \\ 
& \hspace{5cm} 0\leq J \leq \floor{n/2}-1 ~.
\label{eq:HSFSBasisGTrans2}&
\end{align}
\end{subequations}
Note that the generalised binomial coefficient $\binom{n}{k}$ has the following property,
\begin{equation}
\binom{n}{k}=0~~, \qquad k> n~,
\end{equation}
which allows us to take the sum from $J=0$ to $\floor{n}{2}$ for both basis fields in the field strength expression. 
Therefore, the variation of the field strength \eqref{eq:HSFSUniversal} is given by
\begin{align}
2^{n} \delta W_{\a(n)}
=&\hf \sum\limits_{J=0}^{\floor{n/2}}\bigg\{\binom{n}{2J}\delta X^{J}_{\a(n)}+ \frac{1}{2}\binom{n}{2J+1}\delta Z^{J}_{\a(n)}\bigg\}\nonumber &\\
=& \mri \sum\limits_{J=0}^{\floor{n/2}}\bigg\{\binom{n}{2J}\frac{J}{n} \Box^{J}\pde_{(\a_{1}}{}^{\b_{1}}\pde_{\a_{2}}{}^{\b_{2}}...\pde_{\a_{n-2J}}{}^{\b_{n-2J}}D_{\a_{n-2J+1}}\bar{D}^{2}L_{...\a_{n})\b(n-2J)}\nonumber &\\
&-
\binom{n}{2J+1}\frac{n-2J-1}{2n} \non &\\
&\times
\Box^{J+1}\pde_{(\a_{1}}{}^{\b_{1}}\pde_{\a_{2}}{}^{\b_{2}}...\pde_{\a_{n-2J-2}}{}^{\b_{n-2J-2}}D_{\a_{n-2J-1}}\bar{D}^{2}L_{\a_{n-2J}...\a_{n})\b(n-2J-2)}\nonumber &\\
&+\binom{n}{2J}\frac{n-2J}{2n}\Box^{J}\pde_{(\a_{1}}{}^{\b_{1}}\pde_{\a_{2}}{}^{\b_{2}}...\pde_{\a_{n-2J}}{}^{\b_{n-2J}}D_{|\b_{1}}\bar{D}^{2}L_{\b_{2}...\b_{n-2J}|\a_{n-2J+1}...\a_{n})}\nonumber &\\
&-
\binom{n}{2J+1}\frac{2J+1}{2n}\Box^{J}\pde_{(\a_{1}}{}^{\b_{1}}\pde_{\a_{2}}{}^{\b_{2}}...\pde_{\a_{n-2J}}{}^{\b_{n-2J}}D_{|\b_{1}}\bar{D}^{2}L_{\b_{2}...\b_{n-2J}|\a_{n-2J+1}...\a_{n})}\bigg\}\nonumber &\\
=&\sum\limits_{J=0}^{\floor{n/2-1}}\frac{1}{2}\bigg\{\left(\binom{n}{2J+2}\frac{2J+2}{n}-\binom{n}{2J+1}\frac{n-2J-1}{n}\right)\nonumber &\\
&\times \Box^{J+1}\pde_{(\a_{1}}{}^{\b_{1}}\pde_{\a_{2}}{}^{\b_{2}}...\pde_{\a_{n-2J-2}}{}^{\b_{n-2J-2}}D_{\a_{n-2J-1}}\bar{D}^{2}L_{\a_{n-2J}...\a_{n})\b(n-2J-2)}\bigg\}\nonumber &\\
&+\sum\limits_{J=0}^{\floor{n/2}-1}\frac{1}{2}\bigg\{\left(\binom{n}{2J}\frac{n-2J}{n}-\binom{n}{2J+1}\frac{2J+1}{2n}\right)\nonumber &\\
&\times \Box^{J}\pde_{(\a_{1}}{}^{\b_{1}}\pde_{\a_{2}}{}^{\b_{2}}...\pde_{\a_{n-2J}}{}^{\b_{n-2J}}D_{|\b_{1}}\bar{D}^{2}L_{\b_{2}...\b_{n-2J}|\a_{n-2J+1}...\a_{n})}\bigg\}
=0~,
\label{eq:HSFSVar1}&
\end{align}
since for any positive integer $n$ the following identities hold:
\begin{subequations}\label{eq:CoeffConditions}
\begin{align}
\binom{n}{2J+2}\frac{2J+2}{n}-\binom{n}{2J+1}\frac{n-2J-1}{n}= 0 ~,&\quad \forall J 
~;
 &\\
\binom{n}{2J}\frac{n-2J}{n}-\binom{n}{2J+1}\frac{2J+1}{n}=0~,&  \quad \forall J ~. &
\end{align}
\end{subequations}


\subsection{Bianchi identities}

We now prove that 
the field strength \eqref{eq:HSFSUniversal} 
obeys the Bianchi identities \eqref{6.244}
This amounts to computing the following relations:
\begin{subequations}
\begin{align}
D^{\gamma}X^{J}_{\gamma\a(n-1)}=& -\ri \frac{n-2J}{4n}\Box^{J}D^{2}\bar{D}^{\a_{n}}\pde_{\a_{n}}{}^{\b_{n-2J}}\pde_{(\a_{1}}{}^{\b_{1}}...\pde_{\a_{n-2J-1}}{}^{\b_{n-2J-1}}H_{...\a_{n-1})\b(n-2J)}\nonumber &\\
&-\ri\frac{J}{2n}\Box^{J}D^{2}\bar{D}^{\gamma}\pde_{(\a_{1}}{}^{\b_{1}}...\pde_{\a_{n-2J}}{}^{\b_{n-2J}}H_{...\a_{n-1})\gamma\b(n-2J)}~, &\\
D^{\gamma}Z^{J}_{\gamma\a(n-1)}=&+\ri\frac{n-2J-1}{4n}\Box^{J+1}D^{2}\bar{D}^{\gamma}\pde_{(\a_{1}}{}^{\b_{1}}...\pde_{\a_{n-2J-2}}{}^{\b_{n-2J-2}}H_{...\a_{n-1})\gamma\b(n-2J-2)}\nonumber &\\
&+\ri\frac{2J+1}{4n}\Box^{J}D^{2}\bar{D}^{\a_{n}}\pde_{\a_{n}}{}^{\b_{n-2J}}\pde_{(\a_{1}}{}^{\b_{1}}...\pde_{\a_{n-2J-1}}{}^{\b_{n-2J}}H_{...\a_{n-1})\b(n-2J)} ~,&
\end{align}
\end{subequations}
and then using them to evaluate 
\begin{align}
2^{n-1}D^{\b} W_{\a(n-1)\b}=&
\sum\limits_{J=0}^{\floor{n/2}}\binom{n}{2J}D^{\b} X^{J}_{\a(n-1)\b}
+ \sum\limits_{J=0}^{\floor{n/2}}\binom{n}{2J+1}D^{\b} Z^{J}_{\a(n-1)\b}~.
\end{align}
This leads to essentially the same calculation as \eqref{eq:HSFSVar1}, grouping the two independent types of structures that appear and showing that the coefficients of each type of structure vanish. In particular, we arrive again at the relations \eqref{eq:CoeffConditions}.


\begin{footnotesize}

\end{footnotesize}

\end{document}